\documentclass[10pt,aps,prl,nobibnotes,nofootinbib,superscriptaddress,twocolumn,
]{revtex4-2}
\usepackage{graphicx} 	%
\usepackage{dcolumn} 	%
\usepackage{bm}   		%
\usepackage{amssymb}   	%
\usepackage{booktabs}   %
\usepackage{multirow}   %
\usepackage{makecell}   %

\usepackage[separate-uncertainty,separate-uncertainty-units=single,product-units=power,range-units=single,range-phrase=\text{--},retain-explicit-plus=true]{siunitx} %
\usepackage[svgnames]{xcolor}	%
\usepackage[version=4]{mhchem}	%

\usepackage{xr} %
\usepackage[caption=false]{subfig}
\usepackage{lipsum}
\usepackage{xspace} %
\usepackage{silence}
\usepackage[colorlinks=true,citecolor=ForestGreen]{hyperref}
\usepackage[nameinlink,capitalise]{cleveref}

\graphicspath{{figs/}}
\def\Tc{\texorpdfstring{\ensuremath{T_\mathrm{c}}\xspace}{Tc}}
\def\Tcx{\texorpdfstring{\ensuremath{T_{\mathrm{c},x}}\xspace}{Tcx}}
\def\Tcy{\texorpdfstring{\ensuremath{T_{\mathrm{c},y}}\xspace}{Tcy}}
\def\Tczero{\texorpdfstring{\ensuremath{T_{\mathrm{c},0}}\xspace}{Tc0}}
\def\SRO{\texorpdfstring{\ce{Sr2RuO4}\xspace}{Sr2RuO4}}
\def\PZT{\texorpdfstring{\ce{\ce{PbZr_xTi_{1-x}O3}}\xspace}{Sr2RuO4}}

\def\Ag{\ensuremath{{A_\mathrm{1g}}}\xspace}
\def\Agg{\ensuremath{{A_\mathrm{2g}}}\xspace}
\def\Agxy{\ensuremath{{A_\mathrm{1g,1}}}\xspace}
\def\Agz{\ensuremath{{A_\mathrm{1g,2}}}\xspace}
\def\Bg{\ensuremath{{B_\mathrm{1g}}}\xspace}
\def\Bgg{\ensuremath{{B_\mathrm{2g}}}\xspace}
\def\Eg{\ensuremath{{E_\mathrm{g}}}\xspace}
\def\Dfh{\ensuremath{D_\mathrm{4h}}\xspace}
\def\cxyp{\ensuremath{{\frac{c_{11}+c_{12}}{2}}}\xspace}
\def\cxym{\ensuremath{{\frac{c_{11}-c_{12}}{2}}}\xspace}

\def\exx{\ensuremath{\varepsilon_{xx}}\xspace}
\def\sxx{\ensuremath{\sigma_{xx}}\xspace}
\def\sxxdiag{\ensuremath{\sigma_{x'x'}^{110}}\xspace}
\def\exxdiag{\ensuremath{\varepsilon_{x'x'}^{110}}\xspace}
\def\eyy{\ensuremath{\varepsilon_{yy}}\xspace}
\def\ezz{\ensuremath{\varepsilon_{zz}}\xspace}
\def\exy{\ensuremath{\varepsilon_{xy}}\xspace}
\def\sxydiag{\ensuremath{\sigma_{x'y'}^{110}}\xspace}
\def\exydiag{\ensuremath{\varepsilon_{x'y'}^{110}}\xspace}
\def\eyz{\ensuremath{\varepsilon_{yz}}\xspace}
\def\ezx{\ensuremath{\varepsilon_{zx}}\xspace}
\def\eAgxy{\ensuremath{\varepsilon_\Agxy}\xspace}
\def\sAgxy{\ensuremath{\sigma_\Agxy}\xspace}
\def\eAgz{\ensuremath{\varepsilon_\Agz}\xspace}
\def\eBg{\ensuremath{\varepsilon_\Bg}\xspace}
\def\sBg{\ensuremath{\sigma_\Bg}\xspace}
\def\eBgg{\ensuremath{\varepsilon_\Bgg}\xspace}
\def\Vp{\ensuremath{V_\mathrm{piezo}}\xspace}
\def\ph{\ensuremath{p_\mathrm{hydro}}\xspace}

\usepackage[
draft,
]{changes}

\newif\ifsuppinpaper \suppinpapertrue

\ifsuppinpaper
\else
\fi

\begin{document}

\title{Direct evidence for the absence of coupling between shear strain and superconductivity in \SRO}

\author{Giordano Mattoni}%
\email{mattoni@scphys.kyoto-u.ac.jp}
\affiliation{Toyota Riken--Kyoto University Research Center (TRiKUC), Kyoto 606-8501, Japan}

\author{Thomas Johnson}
\affiliation{Toyota Riken--Kyoto University Research Center (TRiKUC), Kyoto 606-8501, Japan}

\author{Atsutoshi Ikeda}
\affiliation{Department of Electronic Science and Engineering, Graduate School of Engineering, Kyoto University, Kyoto 615-8510, Japan}

\author{Shubhankar Paul}
\affiliation{Toyota Riken--Kyoto University Research Center (TRiKUC), Kyoto 606-8501, Japan}
\affiliation{Department of Electronic Science and Engineering, Graduate School of Engineering, Kyoto University, Kyoto 615-8510, Japan}
\affiliation{Department of Physics, Indian Institute of Technology Kanpur, Kanpur 208016, India}

\author{Jake Bobowski}
\affiliation{Irving K. Barber Faculty of Science, University of British Columbia, 3333 University Way, Kelowna, British Columbia, V1V 1V7, Canada}

\author{Manfred Sigrist}
\affiliation{Institute for Theoretical Physics, ETH Z\"urich, 8093 Z\"urich, Switzerland}

\author{Yoshiteru Maeno}
\affiliation{Toyota Riken--Kyoto University Research Center (TRiKUC), Kyoto 606-8501, Japan}

\begin{abstract}
The superconducting symmetry of \SRO has been intensely debated for many years.
A crucial controversy recently emerged between shear-mode ultrasound experiments, which suggest a two-component order parameter, and some uniaxial pressure experiments that suggest a one-component order parameter.
To resolve this controversy, we use a new approach to directly apply three different kinds of shear strain to single crystals of \SRO and investigate the coupling to superconductivity.
After characterising the strain by optical imaging, we observe variations of the transition temperature \Tc smaller than \qty{10}{\milli\kelvin\per\percent} as measured by low-frequency magnetic susceptibility, indicating that shear strain has little to no coupling to superconductivity.
Our results are consistent with a one-component order parameter model, but such a model cannot consistently explain other experimental evidence such as time-reversal symmetry breaking, superconducting domains, and horizontal line nodes, thus calling for alternative interpretations.
\end{abstract}

\date{\today}
\maketitle

\section{Introduction}
Strains in a crystal have well-defined irreducible representations (irreps) under the symmetry of a lattice.
Strains are precisely coupled to the superconducting (SC) order parameter (OP) according to their symmetries, and induce changes in \Tc.
Thus, the application of strains to a superconductor and the observation of the changes in \Tc is a precise method of gaining information on the SC OP.
In addition to isotropic hydrostatic pressure, it has recently become clear that uniaxial strains are a powerful tool to investigate superconducting symmetry, since they can profoundly alter crystal symmetry~\cite{hicks2014strong,steppke2017strong,hicks2025probing}.

The OP of unconventional superconductivity in \SRO~\cite{maeno1994superconductivity} is still unresolved despite 30 years of extensive research efforts~\cite{maeno2024still,maeno2024thirty}.
Spin-singlet-like behaviour was newly established from spin susceptibility measured by nuclear magnetic resonance (NMR)~\cite{pustogow2019constraints,ishida2020reduction,chronister2021evidence} and polarised neutrons~\cite{petsch2020reduction}.
Time-reversal symmetry breaking (TRSB), evidenced by an emerging internal magnetic field  below \Tc~\cite{luke1998time}, has been reported by several groups~\cite{shiroka2012mu, higemoto2014investigation, grinenko2021split}.
Magneto-optic Kerr effect (MOKE) also supports the presence of TRSB in the superconducting state~\cite{xia2006high}.
Anomalous switching behaviour in superconducting junctions~\cite{anwar2021review} point to the presence of SC domains that arise with multi-component OP with one-dimensional (1D) or 2D irreducible representations (irreps).

Uniaxial compression in \SRO enhances \Tc from 1.5 to \qty{3.5}{\kelvin}, and it is associated with the changes in the Fermi surface topology (Lifshitz transition) \cite{steppke2017strong,sunko2019direct}.
A splitting between \Tc and appearance of the TRSB phase was observed by muon spin relaxation ($\mu$SR) under uniaxial strain \cite{grinenko2021split}, pointing to a chiral
superconducting OP.
However, the splitting is not observed in specific heat and elastocaloric-effect measurements under uniaxial strain~\cite{li2021high, li2022elastocaloric}, leaving the issue of TRSB unsettled at present.

Another peculiar behaviour of \SRO is the jump at \Tc of the shear-mode ultrasound velocity related to the elastic modulus $c_{66}$~\cite{benhabib2021ultrasound,ghosh2021thermodynamic}.
This unusual coupling between SC OP and shear strain
implies that the OP has two components: 
either
chiral with 2D irreps (e.g., $d_{yz}\pm id_{zx}$),
nematic with 2D irreps (e.g., $d_{yz}$ only, $d_{zx}$ only, or $d_{yz}\pm d_{zx}$),
or a combination of 1D irreps (e.g., $d + ig$, $s+id$).
Results of uniaxial stress and hydrostatic pressure experiments can be combined to compare with ultrasound experiments \cite{jerzembeck2024tc}, but such comparison leads to both qualitative and quantitative discrepancies~\cite{maeno2024thirty}.
Specifically, a V-shaped kink in \Tc as a function of uniaxial strain along the [110] direction was not observed, in contrast with the expectations from the observed jump in the ultrasound $c_{66}$.
It is crucial to resolve this discrepancy in order to reach a conclusion about the superconducting state of \SRO.  

In this study, we investigate the direct effect of shear strain on the \Tc of \SRO.
The magnitudes of sample strains at low temperatures are measured with optical imaging down to \qty{30}{\kelvin}.
Kinks, linear slopes, and quadratic changes in \Tc are evaluated by applying three different kinds of shear strain and comparing the results with the estimates of the Ehrenfest relations from the jumps and slope changes of the elastic-moduli at \Tc.
We find that \Tc does not change within the experimental precision of $\partial{\Tc} / \partial{\exy} = \qty{+-6}{\milli\kelvin\per\percent}$.
This result seriously challenges the interpretation of SC in \SRO in terms of a two-component OP.

\section{Results}
\subsection{Direct shear-strain quantification at low temperature}

\begin{figure}[tb]
    \includegraphics[page=1,width=89mm]{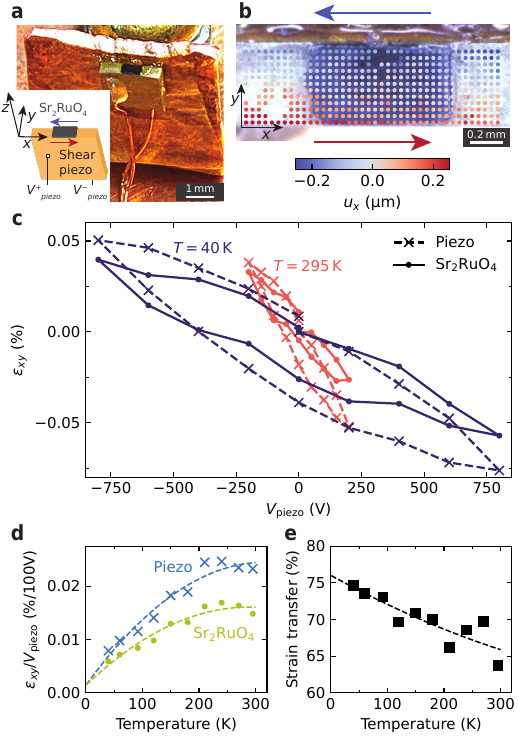}
    \subfloat{\label{fig:SRO_Piezo_Assembly}}
    \subfloat{\label{fig:Imaging_dots}}
    \subfloat{\label{fig:ShearLoops}}
    \subfloat{\label{fig:Shear_vs_Voltage}}
    \subfloat{\label{fig:Strain_transfer}}
    
    \caption{
        \textbf{Optical detection of shear strain.}
        \protect\subref{fig:SRO_Piezo_Assembly} Photo of a thin \SRO crystal (dimensions \qtyproduct{0.8x0.5x0.03}{\milli\metre}, sample S4) attached onto the active side surface of a shear piezo device (dimensions \qtyproduct{2.2x0.5x2.2}{\milli\metre}) and corresponding schematics.
        \protect\subref{fig:Imaging_dots} Optical microscope image of the \SRO sample (black) on the shear piezo surface with \qty{200}{\volt} applied at room temperature.
        The point-by-point horizontal displacement $u_x$ is represented by the coloured dots, while the arrows schematically indicate the in-plane shear strain.
        \protect\subref{fig:ShearLoops} Direct optical measurement of shear strain on the piezo device (crosses) and \SRO (circles) at two selected temperatures.
        \protect\subref{fig:Shear_vs_Voltage} Magnitude of shear at \qty{100}{\volt} and
        \protect\subref{fig:Strain_transfer} amount of shear transferred to the \SRO sample as a function of temperature.
        The dashed lines are polynomial guides to the eye.
    }
    \label{fig:OpticalShear}
\end{figure}

To apply shear strain, we construct piezo-sample assemblies by rigidly gluing thinly polished \SRO crystals directly onto the surface of piezoelectric devices, as shown in \cref{fig:SRO_Piezo_Assembly} (see Methods for construction details).
Deformation of a piezo-sample assembly is monitored by an optical microscope which allows us to calculate the point-by-point in-plane displacements ($u_x$, $u_y$) and the in-plane shear strain
$\exy=(\partial u_x/\partial y+\partial u_y/\partial x)/2$.
We note that this definition is symmetric ($\exy=\varepsilon_{yx}$) and that $\exy=\gamma_{xy}/2=\varepsilon_6/2$, where $\gamma_{xy}$ is the engineering strain and $\varepsilon_6$ is the $xy$ shear strain in Voigt notation.
We show in \cref{fig:Imaging_dots} a typical optical image taken at room temperature with a voltage applied to the shear piezo $\Vp=\qty{+200}{\volt}$.
By means of digital image correlation \cite{schreier2009image,mccormick2010digital}, we extract the point-by-point displacements that are represented by the coloured dots.
As indicated by the arrows, the bottom of the piezo-sample assembly is moving to the right (red) while the top is moving to the left (blue) due to the applied shear strain.
We note that the piezoelectric device we use predominantly generates a shear displacement component $\partial u_x/\partial y$, while the complementary component $\partial u_y/\partial x$ is small due to the device's geometry and boundary constraints (\cref{fig:AllStrain} in the Supplementary Information).
We comment on the importance of measuring both components because relaxation of shear strain could lead to insurgence of a negative $\partial u_y/\partial x$ at the sample surface, which corresponds to an overall sample rotation (\cref{fig:ShearSimulation}).

\begin{figure*}[tb]
    \includegraphics[page=2,width=180mm]{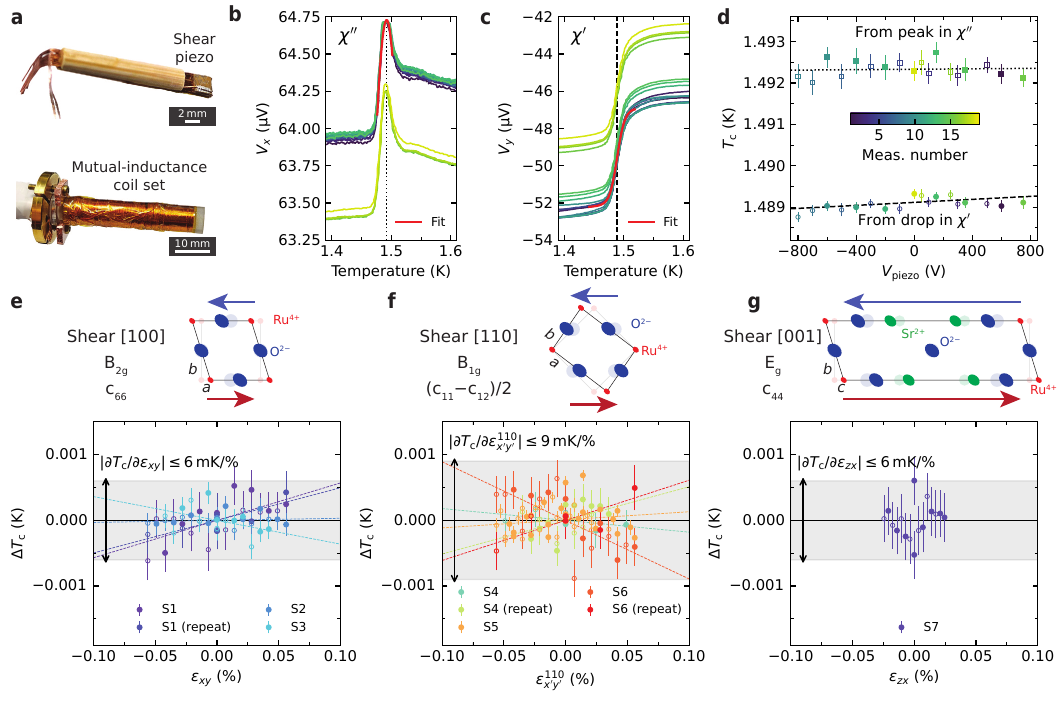}
    \subfloat{\label{fig:ACX_coil}}
    \subfloat{\label{fig:Vx}}
    \subfloat{\label{fig:Vy}}
    \subfloat{\label{fig:dTc_Vshear}}
    \subfloat{\label{fig:dTc_Shear100}}
    \subfloat{\label{fig:dTc_Shear110}}
    \subfloat{\label{fig:dTc_Shear001}}
    
    \caption{
        \textbf{Effect of shear strain on \Tc.}
        \protect\subref{fig:ACX_coil} Assembly of shear piezo and \SRO sample, attached on a bamboo holder with copper wires, to be inserted into a mutual-inductance coil set for measuring AC susceptibility.
        \protect\subref{fig:Vx} Imaginary and
        \protect\subref{fig:Vy} real parts of the AC susceptibility measured by the mutual-inductance method (sample S5).
        The colours represent the chronological order in which the measurements were taken as indicated by the colourscale in \protect\subref{fig:dTc_Vshear}.
        The \Tc is detected by two different criteria: fitting a Lorentzian peak (dotted vertical line in \protect\subref{fig:Vx}) or a sigmoid function (dashed vertical line in \protect\subref{fig:Vy}).
        \protect\subref{fig:dTc_Vshear} Resulting changes in \Tc as a function of voltage applied to the shear piezo: from the peak maximum in $\chi''$ (squares) and from the mid-point of the drop in $\chi'$ (circles).
        We use filled (open) symbols to indicate points taken with increasing (decreasing) voltage.
        \protect\subref{fig:dTc_Shear100} Changes in \Tc with shear strain applied along the [100],
        \protect\subref{fig:dTc_Shear110} [110], and
        \protect\subref{fig:dTc_Shear001} [001] crystalline directions of \SRO using several piezo-sample assemblies.
        The schematics above each panel indicate the corresponding shear deformation of the \SRO lattice.
        A linear regression is performed on each data set to illustrate the variation of the small slopes.
        The grey shaded regions, marked by the black arrows, indicate confidence intervals chosen to represent the scattering of the data of \qty{+-0.6}, \qty{+-0.9}, and \qty{+-0.6}{\milli\kelvin}, respectively.
    }
    \label{fig:ShearTc}
\end{figure*}

We calculate the average shear strain over the surface of the piezo device and the \SRO sample as a function of \Vp in \cref{fig:ShearLoops}.
At room temperature, the piezo device can apply a maximum of about $\exy=\qty{+-0.05}{\percent}$ at $\Vp=\qty{+-200}{\volt}$, corresponding to a shear strain transferred to the \SRO of about $\exy=\qty{+-0.04}{\percent}$.
We emphasize the advantage of this optical technique, which enables non-contact measurement of the effective shear strain directly on the sample surface, allowing us to estimate the shear strain transferred to the bulk of the sample and to neglect the uncertainty due to glue-mediated strain transfer from the piezoelectric device.
The shear--voltage characteristics in \cref{fig:ShearLoops} is nearly linear, with a small hysteresis that is typical of piezoelectric actuators \cite{damjanovic2006hysteresis}.
While this hysteresis is generally negligible for the purposes of this study, it introduces an uncertainty in the shear strain that is most significant around zero applied voltage, where the upper bound estimated from the vertical width of the hysteresis is $\delta\exy\approx\qty{0.02}{\percent}$.
Importantly, this uncertainty is minimal at the largest applied voltages, where the strain response is well defined and reproducible.
As shown in \cref{fig:Shear_vs_Voltage}, the shear applied at a certain voltage becomes gradually smaller at lower temperature due to the decreasing capacitance of the piezoelectric device (\cref{fig:CapacitanceShearConversion}).
We compensate for this effect by applying a larger voltage to the piezo device at low temperature, thus achieving similar values of maximum shear strain at all temperatures.

We show in \cref{fig:Strain_transfer} that the strain transferred from the piezoelectric device to the \SRO sample is as large as \qty{75}{\percent} at lower temperature, where a weak enhancement is possibly caused by a stiffening of the glue used to attach the sample to the piezo surface.
This strain transfer is consistent with the value predicted by our finite element simulations (\cref{fig:ShearSimulation}) for samples of \qty{30}{\micro\metre} thickness.
A direct correlation between the travel distance of the piezoelectric device and its capacitance (\cref{fig:CapacitanceShearConversion}) allows us to calculate a conversion factor $\exy/\Vp = \qty{0.007}{\percent\per 100 \volt}$ at \qty{2}{\kelvin}.
Because the samples investigated in this study had similar thicknesses, we found this conversion factor to be approximately constant.

\subsection{Response of \Tc to shear strain}

We now consider changes in \Tc by shear strain.
To detect the superconducting transition, we insert the piezo-sample assembly of \cref{fig:SRO_Piezo_Assembly} into the mutual-inductance coil set shown in \cref{fig:ACX_coil}.
The mutual-inductance technique allows us to measure the imaginary $\chi''$ and real $\chi'$ parts of the alternating current (AC) susceptibility extracted from the in-phase and out-of-phase lock-in voltages $V_x$, $V_y$, respectively.
These are measured as a function of temperature for different applied voltages \Vp in \cref{fig:Vy,fig:Vx}.
We use \SRO samples with a sharp superconducting transition that appears as a dissipation peak in $\chi ''$, with full width at half maximum of about \qty{20}{\milli\kelvin}, and as a drop in $\chi '$ that indicates the onset of diamagnetic susceptibility, with a transition width smaller than \qty{40}{\milli\kelvin}.
We compare in \cref{fig:dTc_Vshear} two possible criteria for defining \Tc: from the position of the peak in $\chi''$ (squares), and from the mid-point of the drop in $\chi'$ (dots).
We comment that both criteria allow us to overcome time-dependent changes in the electronic signal which suffers from small voltage offsets over time.
The values of \Tc resulting from the two criteria show consistent trends, with a minor difference in absolute value.
Since the peak in $\chi''$ is about an order of magnitude smaller than the drop in $\chi'$, the latter provides a more reliable criterion across different samples and will be used to determine \Tc throughout this work.

Using several piezo-sample assemblies with \SRO glued along different crystalline axes, we apply continuous shear strains along the three directions shown in \cref{fig:dTc_Shear100,fig:dTc_Shear110,fig:dTc_Shear001}:
a pure \exy shear along the [100],
a diagonal shear $\exydiag = \exx -\eyy$ along the [110],
and a $c$-axis shear \ezx along the [001].
As shown in the insets above the panels, these correspond to \Bgg, \Bg, and \Eg strain modes, respectively.
We use several samples from at least two different batches of crystals for the measurements.
We stress that, in contrast to ultrasound experiments where the response of \SRO to shear modes was measured with strains oscillating at high frequencies, our technique allows us to apply static shear strains across the whole dynamic range of the piezoelectric device.
Due to the $C_{4z}$ rotational symmetry of the lattice, a positive and a negative shear are expected to yield identical results [i.e., $\Delta\Tc(\exy) = \Delta\Tc(-\exy)$], hence a symmetric trend is expected around $\exy=0$.
The data in \cref{fig:dTc_Shear100,fig:dTc_Shear110,fig:dTc_Shear001} show no change of \Tc within our experimental resolution for all three kinds of applied shear strain, with neither a kink nor a quadratic trend.
We estimate upper limits of a purely linear slope of the kind $\partial \Tc/\partial\varepsilon$ by drawing grey shaded regions that correspond to \Tc variations smaller than \qty{+-6}, \qty{+-9}, and \qty{+-6}{\milli\kelvin\per\percent}, respectively.
Based on the same shaded regions, upper limits for purely quadratic slopes $\partial^2 \Tc/\partial\varepsilon^2$ are estimated to be \qty{+-60}, \qty{+-90}, and \qty{+-60}{\milli\kelvin\per\percent\squared}.
These values will be further discussed in the following and compared with thermodynamic predictions from ultrasound experiments in \cref{tab:Ehrenfest}.

We now consider the possible effect of thermally-induced pre-strains to \SRO due to the difference in thermal contraction with the piezoelectric device.
As we show in \cref{fig:PreStrains}, an \SRO sample glued along the $ab$ plane is affected by a biaxial tensile pre-strain up to about \qty{+0.1} to \qty{+0.3}{\percent}.
For a [100] piezo-sample assembly, this pre-strain can be seen as a combination of compressive \Ag and shear \Bg strain modes.
In such configuration, the voltage-driven shear-piezo device applies a pure \Bgg strain, a shear mode that does not contain any pre-strain.
Therefore, the \Tc variations that we investigated provide direct information of the SC coupling to a specific shear mode.

\begin{table*}[htb]
\centering
\caption{
\textbf{Comparison between strain and ultrasound.}
The three shear modes investigated in this work are presented along with two compressive modes.
Labels are given for strains, irreducible representations (irreps) in the \Dfh point group, and elastic moduli.
In the middle columns, the inital slope of \Tc change with strains is compared to the estimate from the jump of the elastic moduli at \Tc obtained via the first order Ehrenfest relations \cite{sigrist2002ehrenfest}.
The right columns compare the second derivative of \Tc versus strain with the slope change of the elastic moduli at \Tc calculated via the second order Ehrenfest relations.
}
\label{tab:Ehrenfest}
\resizebox{\textwidth}{!}{
\begin{tabular}{lccc|ccc|ccc}

 ~ &
 ~ &
 ~ &
 ~ &
 \multicolumn{3}{c|}{\textbf{$1^\mathrm{st}$ order} (jump at \Tc)} &
 \multicolumn{3}{c}{\textbf{$2^\mathrm{nd}$ order} (slope change at \Tc)} \\

 ~ &
 \textbf{Strain} &
 \textbf{Irrep} &
 \textbf{Mod.} &
 \textbf{Strain} &
 \textbf{Ultras.} &
 \textbf{Ehrenfest rel.} &
 \textbf{Strain} &
 \textbf{Ultras.} &
 \textbf{Ehrenfest rel.}\\

~ &
~ &
~ &
~ &
(\qty{}{\milli\kelvin\per\percent}) &
(\qty{}{\milli\kelvin\per\percent}) &
~ &
(\qty{}{\milli\kelvin\per\percent\squared}) &
(\qty{}{\milli\kelvin\per\percent\squared}) &
~ \\

\midrule

\multirow[c]{3}{*}{\rotatebox[origin=c]{90}{Shear}}&
\exy &
\Bgg &
$c_{66}$ &
$<\qty{6}{}$ &
70~\cite{benhabib2021ultrasound}, 750~\cite{ghosh2021thermodynamic} &
$\displaystyle\left|\frac{\partial\Tc}{\partial\exy}\right|\sim\sqrt{-2\frac{\Tc \Delta c_{66}}{\Delta C}}$ &
$<\qty{60}{}$  &
280~\cite{ghosh2021thermodynamic} &
$\displaystyle\frac{\partial^2\Tc}{\partial\exy^2}=-\frac{\Tc }{\Delta C}\Delta\frac{\partial c_{66}}{\partial T}$ \\[4mm]

~ &
\makecell{\exydiag \\ $\exx - \eyy$} &
\Bg &
\cxym &
$<\qty{9}{}$ &
not detected &
$\displaystyle\left|\frac{\partial\Tc}{\partial\varepsilon_\Bg}\right|\sim\sqrt{-\frac{1}{2}\frac{\Tc \Delta c_\Bg}{\Delta C}}$ &
$<\qty{90}{}$ &
$-$7300~\cite{ghosh2021thermodynamic} &
$\displaystyle\frac{\partial^2\Tc}{\partial\varepsilon_\Bg^2} =-\frac{\Tc }{\Delta C}\Delta\frac{\partial c_\Bg}{\partial T}$ \\[4mm]

~ &
\eyz, \ezx &
\Eg &
$c_{44}$ &
$<\qty{6}{}$ &
--- &
no coupling &
$<\qty{60}{}$ &
--- &
no coupling \\

\midrule

\multirow[c]{2}{*}{\rotatebox[origin=c]{90}{Compr.}} &
$\exx + \eyy$ &
\Agxy &
\cxyp &
320 \cite{maeno2024thirty} &
1600~\cite{ghosh2021thermodynamic} &
$\displaystyle\left|\frac{\partial\Tc}{\partial\varepsilon_\Agxy}\right|\sim\sqrt{-\frac{\Tc \Delta c_\Agxy}{\Delta C}}$ &
--- &
2500~\cite{ghosh2021thermodynamic} &
$\displaystyle\frac{\partial^2\Tc}{\partial\varepsilon_\Agxy^2}=-\frac{\Tc }{\Delta C}\Delta\frac{\partial c_\Agxy}{\partial T}$ \\[4mm]

~ &
\ezz &
\Agz &
$c_{33}$ &
310 \cite{maeno2024thirty} &
1600~\cite{ghosh2021thermodynamic} &
$\displaystyle\left|\frac{\partial\Tc}{\partial\ezz}\right|\sim\sqrt{-\frac{\Tc \Delta c_{33}}{\Delta C}}$ &
--- &
--- &
--- \\[4mm]

\end{tabular}
}
\end{table*}

\subsection{Tensile control experiment}
\begin{figure}[tb]
    \includegraphics[page=3,width=89mm]{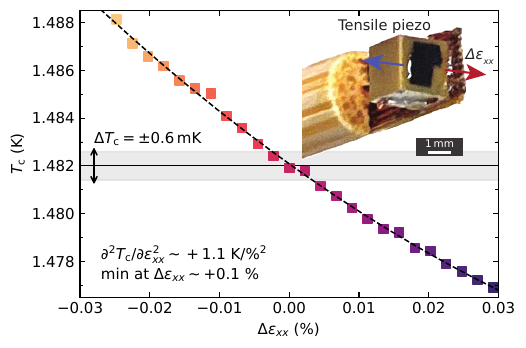}
    \caption{
        \textbf{Tensile strain applied to \SRO.}
        Variations of \Tc measured by mutual inductance with an assembly of a tensile piezoelectric device and an \SRO sample.
        A quadratic dependence on uniaxial strain is observed, with a minimum at $\Delta\exx\sim\qty{+0.1}{\percent}$ that indicates presence of pre-strains.
        To compare the \Tc resolution with the shear data, the grey shaded region indicates an interval of \qty{+-0.6}{\milli\kelvin}.
    }
    \label{fig:TensileTc}
\end{figure}

To validate our technique of detecting small \Tc changes with strain, we perform a similar experiment with a uniaxial tensile piezo.
As shown in \cref{fig:TensileTc}, tensile strain applied to \SRO induces a clear \Tc change in the range of several \qty{}{\milli\kelvin} (see \cref{fig:TensileCalibration} for a voltage-to-strain conversion and \cref{fig:TensileExtended} for extended data).
As indicated by the fitted curve (black dashed line), the strain-dependence of \Tc follows a quadratic behaviour.
This quadratic trend is consistent with that reported by Watson \textit{et al.} \cite{watson2018micron}, although our minimum is not at $\Delta\exx=0$.
A likely cause of this shift is the presence of pre-strains caused by the differential thermal contraction of \SRO and the piezoelectric device~\cite{shapiro2016measurement}.
We discuss in \cref{fig:PreStrains} that for an assembly of \SRO glued over the active surface of a tensile piezo there is a biaxial pre-strain up to about \qty{+0.1} to \qty{+0.3}{\percent}, which is reasonably consistent with the value of \qty{+0.1}{\percent} extracted from the fit in \cref{fig:TensileTc}.
Although it is difficult to extract quantitative results from this control experiment with tensile strain, it proves that the technique used in this work is suitable to capture the smallest variations of \Tc.

\section{Discussion}
Our results show that the \Tc of \SRO does not change under the direct application of shear strains \exy, \exydiag, and \ezx.  
This is different from the \Tc variations that occur under compressive strains.
In the following, we compare our results with the prediction of shear-strain response estimated by combining results of uniaxial compressions and hydrostatic pressure.
We also examine the limit of compatibility with the jumps in ultrasound velocities at \Tc and with the change in their slope.
Finally, we compare with the jumps in thermal expansion at \Tc.

\subsection{Comparison with uniaxial stress and ultrasound experiments}

\begin{figure}[!b]
    \includegraphics[page=4,width=75mm]{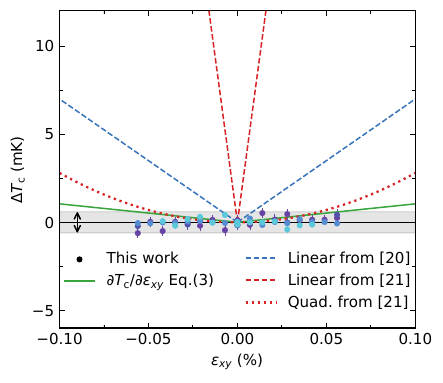}
    \caption{
        \textbf{Comparison with other stress--strain and ultrasound experiments.}
        Our experimental data from \cref{fig:dTc_Shear100} is shown as circles, while the grey shaded region marked by the black arrows indicates our confidence interval for the lack of a kink or a quadratic strain dependence.
        The green solid line shows the kink required by the symmetry with the slope estimated combining stress experiments from Refs. \cite{jerzembeck2024tc,forsythe2002evolution,jerzembeck2022superconductivity} via \cref{eq:xy_from_ThreeExperiments}.
        The V-shaped kinks shown by the blue and red dashed lines are estimated from the $c_{66}$ ultrasound jump at \Tc~\cite{benhabib2021ultrasound,ghosh2021thermodynamic} via the 1\textsuperscript{st} order Ehrenfest relations.
        The dotted red line shows the quadratic slope calculated via the 2\textsuperscript{nd} order Ehrenfest relation from the $c_{66}$ slope change at \Tc~\cite{ghosh2021thermodynamic}.
        }
    \label{fig:StrainUltrasound}
\end{figure}

In \cref{tab:Ehrenfest} (left columns), we give some of the irreducible representations (irrep) of the \Dfh point group that apply to space group I4/mmm of \SRO crystal structure.
The couplings between different strains and superconductivity are dictated by symmetry.
Due to the $C_{4z}$ rotational symmetry, the \Tc variation under shear is expected to be symmetric and show either a linear kink for a two-component SC or a quadratic dependence for a one-component SC.
For the two-component order parameter $\bm{\eta} = (\eta_x, \eta_y)$ that belongs to \Eg irrep with components $\{d_{xz}, d_{yz}\}$, the Ginzburg-Landau (GL) free energy density that describes the coupling to lattice strain is given by (more details in the Supplementary Information)
\begin{equation}
    \label{eq:GLenergy2D-OP}
    \begin{aligned}
        f_{\eta \varepsilon} &= r_1 \eAgxy|\bm{\eta}|^2 + r_2 \eAgz |\bm{\eta}|^2 \\
        &+ r_3 \eBgg (\eta_x^* \eta_y + \eta_x \eta_y^*) + r_4 \eBg (|\eta_x|^2 - |\eta_y|^2),
    \end{aligned}
\end{equation}
up to second order in {$\eta_x, \eta_y$} and first order in strain, with constant coupling parameters $r_i$~\cite{sigrist2002ehrenfest}.
The \Dfh symmetry does not allow linear coupling to strains \eyz and \ezx of irrep \Eg and modulus $c_{44}$, so to lowest order these strains should not affect \Tc.
For a nematic or chiral OP, we have
\begin{equation}
    \begin{aligned}
    \frac{\partial \Tc}{\partial \eBgg} &= \pm \left|r_3\right|\:\: (\pm\ \mathrm{sign\ for}\ \eBgg \gtrless 0),
    \\
    \frac{\partial \Tc}{\partial \eBg} &= \pm \left|r_4\right|\:\: (\pm\ \mathrm{sign\ for}\ \eBg \gtrless 0),
    \end{aligned}
\end{equation}
indicating that application of pure shear strains \Bgg or \Bg is expected to lead to a kink in \Tc around $\varepsilon=0$.
We show again with a grey confidence interval in \cref{fig:StrainUltrasound} that our experiment proves little to no coupling to \exy.

We now compare our direct measurement of shear strain with estimates from other techniques where shear strain is indirectly applied along other strains.
By combining the results of other stress--strain experiments (details in the Supplementary Information and \cref{tab:StressStrain}) as
\begin{align}
\label{eq:xy_from_ThreeExperiments}
\frac{\partial \Tc}{\partial \exy} =
2c_{66}\left(
2\frac{d\Tc}{d\sxxdiag}
+\frac{d\Tc}{d\ph}
+\frac{\partial \Tc}{\partial \sigma_{zz}}
\right),
\end{align}
where \sxxdiag is the uniaxial compression along the [110] direction,
$\sigma_{zz}$ is uniaxial compression along the $z$ axis,
and the hydrostatic pressure \ph is positive for compression, (opposite sign compared to the $\sigma_i$).
By using
$c_{66}=\qty{65.5}{\giga\pascal}$ \cite{ghosh2021thermodynamic}
and values of the three slopes
\qty{+71}{\milli\kelvin\per\giga\pascal} \cite{jerzembeck2024tc},
\qty{-210}{\milli\kelvin\per\giga\pascal} \cite{forsythe2002evolution}, and
\qty{+76}{\milli\kelvin\per\giga\pascal} \cite{jerzembeck2022superconductivity},
respectively,
we obtain $\partial \Tc/\partial \exy \sim \qty{10.5}{\milli\kelvin\per\percent}$, which is shown by the green V-shaped line in \cref{fig:StrainUltrasound}.
This estimate is consistent with the one provided by by Jerzembeck \textit{et al.}
$\left|{\partial \Tc}/{\partial \exy}\right| < \qty{13}{\milli\kelvin\per\percent}$~\cite{jerzembeck2024tc}.

Interestingly, we observe no measurable change in \Tc under \exydiag, which corresponds to pure \Bg shear strain \cref{fig:dTc_Shear110}.
This is in contrast to previous uniaxial stress studies along [100], where an increase in \Tc was reported and attributed to the \Bg symmetry channel \cite{barber2019role}.
In those works, the strain state includes a combination of symmetry components due to the Poisson effect, including a significant \Ag contribution.
By contrast, our setup applies a pure 
\Bg shear, with negligible \Ag component.
Furthermore, the strain magnitude in our experiment is smaller, such that the Fermi surface remains far from the van Hove singularity.
At present, the absence of a \Tc response under pure \Bg strain remains an open question, but it highlights the complexity of how different strain symmetries couple to the superconducting state.

In addition, we compare our results with estimates from 
the thermodynamic Ehrenfest relations
\begin{equation}
   \label{eq:Ehrenfest}
   \left| \frac{\partial \Tc}{\partial \varepsilon_{ij}} \right| = \sqrt{- \frac{\Tc \Delta c_{ij}}{f(\beta_k) \Delta C}},
\end{equation}
where $\Delta c_{ij}$ is the jump in elastic moduli at \Tc,
$\Delta C\sim\qty{570}{\joule\per\cubic\meter\per\kelvin}$ \cite{ghosh2021thermodynamic} is the jump in specific heat and $f(\beta_k)$ is a function of the 4\textsuperscript{th} order GL free energy coefficients $\beta_k$~\cite{sigrist2002ehrenfest}.
Here, the coefficients $f(\beta_k)$ are parameters that can be estimated in the weak-coupling limit under the isotropic-band approximation~\cite{sigrist2002ehrenfest}, yielding the equations and estimates presented in the central columns of \cref{tab:Ehrenfest}.
As shown in \cref{fig:StrainUltrasound}, the V-shaped kinks estimated from Ehrenfest relations have a much larger slope than our direct measurement.
It should be noted that discrepancies are also serious for the \Ag compressional modes $(c_{11}+c_{12})/2$ and $c_{33}$: the \Tc variations under constant stress are both about a factor five smaller than the expectation from the compressional-mode ultrasound jumps~\cite{maeno2024thirty}. 
Thus, we find here an inconsistency with the Ehrenfest relations, which has to be resolved.

Finally, we consider the upper limit for quadratic variation in \Tc found in this study (grey regions of \cref{fig:dTc_Shear100,fig:dTc_Shear110,fig:dTc_Shear001},) and compare it with estimates from second order Ehrenfest relations that depend on the slope jump of elastic moduli at \Tc (right columns of \cref{tab:Ehrenfest}) \cite{sigrist2002ehrenfest}.
While the quadratic term estimated from $c_{66}$ is rather small,
the one from $(c_{11}-c_{12})/2$ is very large.
It is rather puzzling that a second-order coupling is so large in the absence of the first-order coupling.
Such discrepancy in both shear and compressional modes suggests that the ultrasound response may contain additional contributions not given by the coupling to the SC OP.
In the normal state of \SRO, the elastic moduli $c_{11}$ and $(c_{11}-c_{12})/2$ are known to exhibit a peculiar temperature dependence compared to the other modes~\cite{okuda2002unconventional}.
These modes involve stretching and shrinking of the in-plane Ru-O bond lengths that may cause additional coupling to the quasiparticles, although it is not clear how such coupling changes below \Tc.

\subsection{Comparison with thermal expansion}
Another thermodynamic comparison can be made with jumps in thermal expansion coefficient $\alpha_{i} = \partial\varepsilon _{ii} / \partial T$ using the different Ehrenfest relation

\begin{equation}
   \label{eq:thermal expansion Ehrenfest}
   \frac{\partial \Tc}{\partial \sigma_{ii}} 
   = - \Tc \frac{\Delta \alpha _i}{\Delta C},
\end{equation}
where $C$ is the specific heat, and $\Delta$ refers to the jumps at \Tc.
Note that, unlike \cref{eq:Ehrenfest} for elastic moduli, which involves elastic couplings to the superconducting order-parameter in the free energy, this Ehrenfest relation is a more direct thermodynamic requirement.   
From the negative jumps
$\Delta \alpha_{a} = \qty{-0.42e-7}{\per\kelvin}$ and
$\Delta \alpha_{c} = \qty{-0.52e-7}{\per\kelvin}$
observed by Grube \textit{et al.}~\cite{grube2024anomalous}, we obtain
$\partial \Tc/\partial \ph = \qty{-360}{\milli\kelvin\per\giga\pascal}$ and
$\partial \Tc/\partial \sigma_{zz} = \qty{+140}{\milli\kelvin\per\giga\pascal}$.
These estimates are within a factor two in agreement with the results of
hydrostatic pressure \qty{-210}{\milli\kelvin\per\giga\pascal}~\cite{forsythe2002evolution} and
$c$-axis uniaxial-stress \qty{+76}{\milli\kelvin\per\giga\pascal}~\cite{jerzembeck2022superconductivity}.
Thus, the strong disagreements of ultrasound jumps in both shear and compression modes suggest that the ultrasound jumps may contain additional contributions which are not properly included in the Ehrenfest relation of \cref{eq:Ehrenfest}.

\section{Conclusions}
We investigated the coupling of \SRO superconductivity to shear strain.
For this purpose, we developed a technique where thin sample crystals were glued on shear-strain piezoelectric devices that allowed us to directly apply static shear strains along three symmetry channels.
The magnitudes of strains were evaluated from optical imaging down to \qty{30}{\kelvin}.
We found no detectable change of \Tc, with a resolution limit smaller than \qty{10}{\milli\kelvin\per\percent}. 
This poses strict limits on the possible nature of the superconducting order parameter, pointing at ruling out a two-component order parameter.
We note, however, that there remains an unresolved discrepancy between the results of ultrasound and static deformation (stress/strain) experiments, which requires further investigation.
Our analysis indicates that \SRO has a one-component OP, or it is a host of more exotic OPs that are required to fully explain its superconducting state.
The new method developed in this study can be readily applied to other superconductors for which a two-component OP is suggested, such as \ce{UPt3}, or to other materials showing different kinds of phase transitions where a two-component order parameter may emerge.

\section{Methods}

\subsection{Piezo-sample assemblies and sample crystals}
In order to apply various shear strains to \SRO samples, we used a shear piezoelectric device (Thorlabs, PL5FB) consisting of a single piezoelectric ceramic layer of \PZT (PZT) with gold-plated electrodes on the top and bottom surfaces.
The device provides a free-stroke displacement of \qty{1.3}{\micro\metre} with the application of \qty{+- 200}{\volt} at room temperature.
We cut the commercial \qtyproduct{5x5x0.5}{\milli\metre} shear devices into four pieces of about \qtyproduct{2x2x0.5} {\milli\metre} with a diamond-disc cutter, and polished the cut planes using a polishing machine (Musashino Denshi, MA-200) with a diamond slurry of \qty{3}{\micro\metre}.
We left a suitable surface roughness to secure strong adhesion to the sample crystal and to provide a clear optical contrast for measuring actual strain.
For the control experiment, we used a tensile strain device (Thorlabs, PA3CE, dimension \qtyproduct{2x2x2}{\milli\metre}) consisting of piezoelectric layers stacked in series.
The device provides a free-stroke expansion displacement of \qty{2}{\micro\metre} with the application of \qty{+100}{\volt} at room temperature.

The \SRO single crystals used in this study were grown by a floating-zone method using an image furnace (Canon Machinery, SC-E15HD)~\cite{bobowski2019improved}.
The cystal orientation was determined by X-ray Laue patterns (Rigaku, RASCO BL-II with a CCD camera).
We selected \SRO crystals with a sharp superconducting transition at $\Tc = \qty{1.5}{\kelvin}$, ensuring the absence of metallic-\ce{Ru} inclusions that would induce broad enhancement of \Tc up to \qty{3.5}{\kelvin} \cite{maeno1998enhancement}.
The \SRO crystals were cut into a typical size of \qtyproduct{0.4x1}{\milli\metre} and polished down to a thickness of \qty{30+-5}{\micro\metre} with a diamond slurry of \qty{3}{\micro\metre}.
Each thin crystalline sample was glued to the active side surface of a piezo device with epoxy (EPO-TEK, 353ND) cured at \qtyrange{110}{120}{\degreeCelsius} for \qty{20}{\minute}.
The lap shear of EPO-TEK 353ND (\qty{13.5}{\mega\pascal})
is much larger than the typical lap shear of other low-temperature epoxy such as Stycast 2850FT (\qty{3}{\mega\pascal}, with catalyst 24LV),
ensuring good adhesion between the \SRO and the piezo element.

High-voltage electrodes for the piezo elements were provided by copper wires (diameter \qty{0.14}{\milli\metre}) attached using silver paste (EPO-TEK, H20E) cured at \qty{120}{\degreeCelsius} for \qty{2}{\hour}.
For SC measurements, each piezo-sample assembly was mounted on a bamboo holder on which ten insulated copper wires (diameter \qty{0.1}{\milli\metre}) were used for thermal anchoring, glued with Varnish (GE, 8530) as shown in \cref{fig:ACX_coil}.

\subsection{Image correlation to optically measure the strains}
We employed an optical-imaging technique to quantify the shear and tensile strain applied by the piezoelectric devices to \SRO samples.
Digital photographs were taken with different \Vp using an optical microscope (MicroSupport, SS306) with \qtyrange{10}{40}{\times} zoom.
Digital image correlation (DIC) with DICe software was employed to compute in-plane displacements in the $x$ and $y$ directions \cite{schreier2009image,mccormick2010digital}.
The average strain over the sample and piezo surfaces was calculated line-by-line averaging the displacement data, while the error was estimated from the standard deviation of the data.

\subsection{Calibration of strain}
The conversion factor from \Vp to shear \exy and uniaxial \exx was measured at different temperatures between 30 and \qty{295}{\kelvin} in an optical cryostat (Thermalblock, custom design).
Since the piezo displacement per unit volt depends on temperature, the capacitance of the piezo was used to bridge between the measurements at \qty{30}{\kelvin}, lowest temperature accessible by our optical cryostat, and the measurements of \Tc which were performed at and below \qty{2}{\kelvin}.
The strain transfer ratio discussed in \cref{fig:Strain_transfer} was estimated on different samples and found to be rather constant among the several samples due to their comparable thicknesses.

\subsection{Determination of \Tc}
Evaluation of \Tc with exceptionally high precision and accuracy was crucial in this study.
We adopted alternating-current (AC) susceptibility measurements using a mutual-inductance coil set~\cite{yonezawa2015compact} and a lock-in amplifier (Stanford Research Systems, SR830) to probe both the imaginary $\chi''$ and real $\chi'$ parts of the AC susceptibility defined as $\chi = \partial M / \partial H = -i\chi'' + \chi '$.
By driving an alternating current of the form
$I_\mathrm{AC}(t) = I_0 \cos(2\pi f t)$
in the excitation coil, the resulting voltage in the pickup coil is
$V_\mathrm{AC}(t) = (1/\alpha)  V_\mathrm{sample}  I_0  f  (-\chi'' \cos(2\pi f t) + \chi' \sin(2\pi f t))$,
with $\alpha$ a constant related to the geometry of the coils, $V_\mathrm{sample}$ the sample volume, $I_0$ the amplitude of the excitation current, and $f$ its frequency \cite{nikolo1995superconductivity}.
The coil set consisted of a pair of in-series pick-up/reference coils
(400 turns each, internal diameter \qty{5}{\milli\metre}, length \qty{5.5}{\milli\metre}, copper wires diameter \qty{50}{\micro\metre})
wound in opposite directions,
and an excitation coil
(\qty{2400}{turns}, internal diameter \qty{6.0}{\milli\metre}, length \qty{36}{\milli\metre}, copper wires diameter \qty{100}{\micro\metre}),
producing a magnetic field of \qty{84}{\micro\tesla\per\milli\ampere}.
An alternating current (frequency \qty{17.7}{\kilo\hertz}, root mean squared amplitude of \qty{1}{\milli\ampere}) was sourced to the excitation coil using the voltage output of the lock-in amplifier and a series resistance of \qty{10}{\kilo\ohm}.
In-phase ($V_x$) and out-of-phase ($V_y$) lock-in components were calibrated at \qty{2}{K} by considering the phase delay of the circuit involving only the excitation coil.
Several values of current were tested to ensure that the effect of heating was below \qty{1}{\milli\kelvin} near \Tc.
The sample temperature was measured with a thermometer (Lake Shore, Cernox CX-1030-SD-HT) mounted on the copper plate to which each piezo-sample assembly and mutual-inductance coil set were thermally anchored.

\section*{Data availability}
The data that supports the findings of this study is available from the corresponding author upon reasonable request.

\section*{Acknowledgements}
The authors thank K. Grube, S. Kittaka, H. v. L\"ohneysen, and A. Ramires for useful information,
H. Matsuki for his technical contribution,
and N. Manca for comments on the manuscript.
This work was supported by JSPS Grant-in-Aids KAKENHI Nos.
JP26247060, %
JP15H05852, %
JP15K21717, %
JP17H06136, %
JP18K04715, %
22H01168, %
23K22439, %
as well as by JSPS Core-to-Core program.
G.M. acknowledges support from
the Kyoto University Foundation, %
Kakenhi 25K17346, %
and Toyota Riken Scholar. %
T.J. acknowledges support as JSPS International Research Fellow (PE24047).
S.P. is supported by the JST Sakura Science Exchange Program.

\section*{Author contributions}
Y.M. conceived and supervised the project;
T.J, S.P., J.B., and Y.M. prepared the \SRO crystals;
Y.M. prepared the sample-piezo assemblies;
G.M. developed the imaging apparatus, performed the low-temperature measurements, and analysed the data;
T.J., G.M., M.S., and A.I. discussed the theoretical modelling and comparison with the ultrasound data.
G.M., T.J., and Y.M. wrote the manuscript with input from all co-authors.

\ifsuppinpaper

\onecolumngrid
\appendix
\clearpage
\noindent\rule{\textwidth}{1pt}
\begin{center}
	{\huge \texttt{Supplementary Information}}
\end{center}
\addcontentsline{toc}{section}{Supplementary Information}
\noindent\rule{\textwidth}{1pt}

\else

\documentclass[aps,preprint,nobibnotes,superscriptaddress]{revtex4-1}

\externaldocument{ShearSRO}
\begin{document}
	\noindent\rule{\textwidth}{1pt}
	\begin{center}
		{\huge \texttt{Supplementary Information}}
	\end{center}
	\addcontentsline{toc}{section}{Supplementary Information}
	\noindent\rule{\textwidth}{1pt}
	\title{Direct evidence for the absence of coupling between shear strain and superconductivity in \SRO}

\author{Giordano Mattoni}%
\email{mattoni@scphys.kyoto-u.ac.jp}
\affiliation{Toyota Riken--Kyoto University Research Center (TRiKUC), Kyoto 606-8501, Japan}

\author{Thomas Johnson}
\affiliation{Toyota Riken--Kyoto University Research Center (TRiKUC), Kyoto 606-8501, Japan}

\author{Atsutoshi Ikeda}
\affiliation{Department of Electronic Science and Engineering, Graduate School of Engineering, Kyoto University, Kyoto 615-8510, Japan}

\author{Shubhankar Paul}
\affiliation{Toyota Riken--Kyoto University Research Center (TRiKUC), Kyoto 606-8501, Japan}
\affiliation{Department of Electronic Science and Engineering, Graduate School of Engineering, Kyoto University, Kyoto 615-8510, Japan}
\affiliation{Department of Physics, Indian Institute of Technology Kanpur, Kanpur 208016, India}

\author{Jake Bobowski}
\affiliation{Irving K. Barber Faculty of Science, University of British Columbia, 3333 University Way, Kelowna, British Columbia, V1V 1V7, Canada}

\author{Manfred Sigrist}
\affiliation{Institute for Theoretical Physics, ETH Z\"urich, 8093 Z\"urich, Switzerland}

\author{Yoshiteru Maeno}
\affiliation{Toyota Riken--Kyoto University Research Center (TRiKUC), Kyoto 606-8501, Japan}

	\maketitle
	
	\fi
	
	\renewcommand\thefigure{S\arabic{figure}}    
	\setcounter{figure}{0}
	\renewcommand\thetable{S\arabic{table}}    
	\setcounter{table}{0}
	\renewcommand\theequation{S\arabic{equation}}    
	\setcounter{equation}{0}

	\begin{figure*}[htb]
		\includegraphics[page=5,width=170mm]{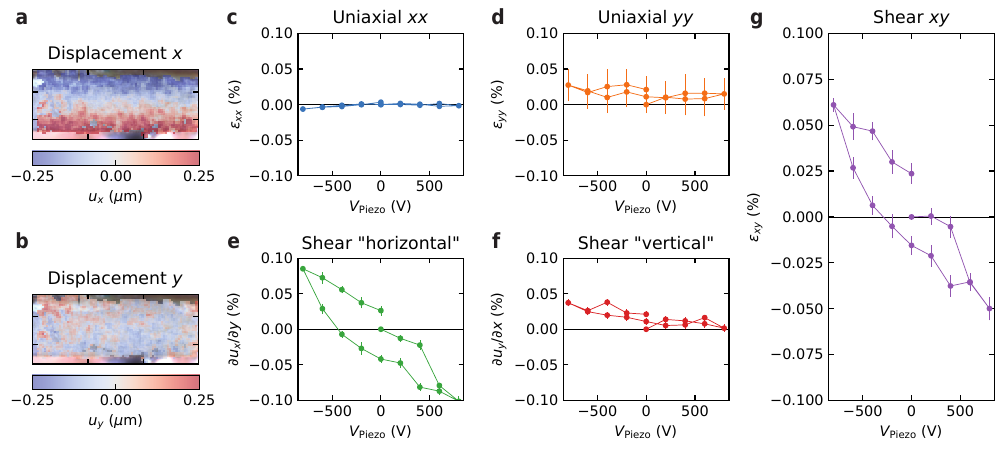}
		\subfloat{\label{fig:AllStrain_dx}}
		\subfloat{\label{fig:AllStrain_dy}}
		\subfloat{\label{fig:AllStrain_exx}}
		\subfloat{\label{fig:AllStrain_eyy}}
		\subfloat{\label{fig:AllStrain_ux}}
		\subfloat{\label{fig:AllStrain_uy}}
		\subfloat{\label{fig:AllStrain_exy}}
		\caption{
			\textbf{All components of in-plane strain applied by a shear-piezo device at \qty{60}{\kelvin}.}
			\protect\subref{fig:AllStrain_dx} Free-stroke displacement along the $x$ and
			\protect\subref{fig:AllStrain_dy} $y$ directions as indicated by the colour scales.
			\protect\subref{fig:AllStrain_exx} Uniaxial strain along the $x$ and
			\protect\subref{fig:AllStrain_eyy} $y$ directions are both negligible.
			\protect\subref{fig:AllStrain_ux} Displacement along the $x$ direction is the predominant shear deformation applied by the piezo, while
			\protect\subref{fig:AllStrain_uy} the deformation along the $y$ direction is negligible.
			\protect\subref{fig:AllStrain_exy} Resulting shear tensor component $\exy = (\partial u_x/\partial y + \partial u_y/\partial x)/2$.
		}
		\label{fig:AllStrain}
	\end{figure*}

	\begin{figure*}[htb]
		\includegraphics[page=6,width=170mm]{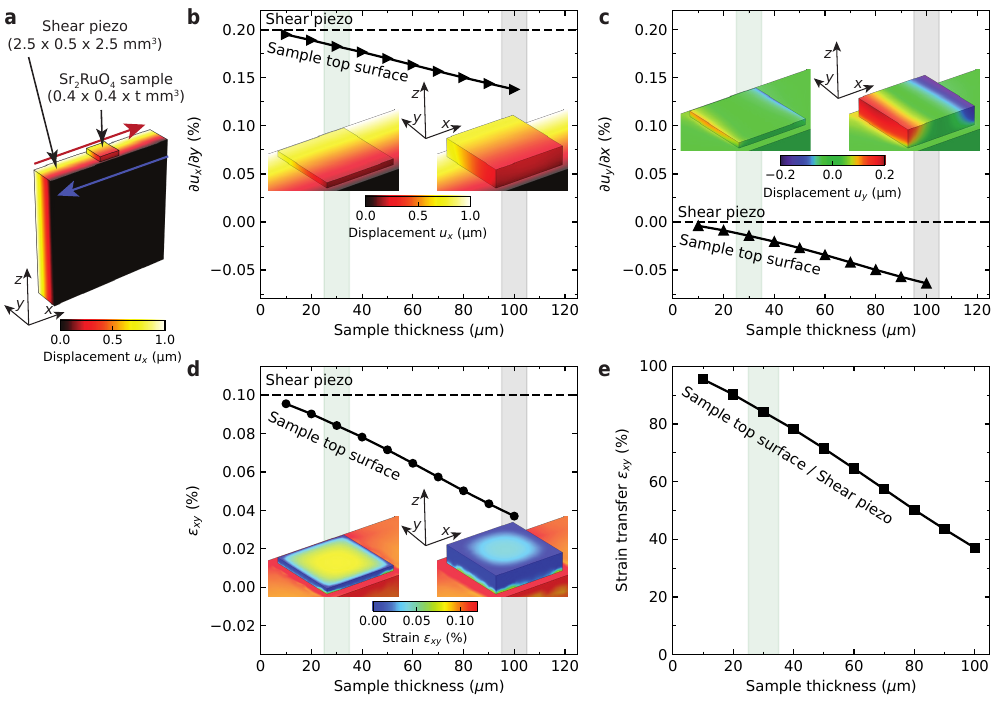}
		\subfloat{\label{fig:ShearSimulation_img}}
		\subfloat{\label{fig:ShearSimulation_ux}}
		\subfloat{\label{fig:ShearSimulation_uy}}
		\subfloat{\label{fig:ShearSimulation_exy}}
		\subfloat{\label{fig:ShearSimulation_transfer}}
		
		\caption{
			\textbf{Finite element simulation of shear strain for different sample thicknesses.}
			\protect\subref{fig:ShearSimulation_img} We simulate an \SRO sample (dimensions \qtyproduct{0.4x0.4}{\milli\metre}, thicknesses between \qtyrange{10}{100}{\micro\metre}) rigidly connected to the active surface of a shear piezo device (dimensions \qtyproduct{2.5x0.5x2.5}{\milli\metre}).
			For simplicity, we neglect any interface effect related to the glue that is used in the experiment to connect the two elements.
			We apply a shear strain $\exy=\qty{0.1}{\percent}$ by displacing the rear $xz$ face of the piezo device by \qty{1}{\micro\metre} in the $x$ direction (yellow colour) while the $xz$ face in the foreground is kept fixed (black colour).
			\protect\subref{fig:ShearSimulation_ux} The shear piezo transfers the displacement $u_x$ to the sample, but this is gradually reduced as the sample thickness increases.
			The insets show the simulated displacement data for 30 and \qty{100}{\micro\metre} sample thickness, while the graph show the strain $\partial u_x/\partial y$.
			\protect\subref{fig:ShearSimulation_uy} While the shear piezo does not apply displacement $u_y$, such displacement naturally emerges in the \SRO sample, and it is bigger for larger sample thicknesses.
			The emerging strain $\partial u_y/\partial x$ has opposite sign of the strain $\partial u_x/\partial y$ considered in the previous panel, indicating an emerging sample rotation.
			Such rotation is the natural outcome of shear strain relaxation along the sample thickness.
			\protect\subref{fig:ShearSimulation_exy} Resulting shear $\exy=1/2( \partial u_x/\partial y + \partial u_y/\partial x )$.
			On the \SRO sample surface, \exy is rather large for small sample thicknesses, while it decreases for thicker samples due to strain relaxation.
			In addition, the sample edges show lower values of \exy.
			In our experiments, we partially counteracted this effect by placing more epoxy glue along the edges of the samples.
			\protect\subref{fig:ShearSimulation_transfer} Transfer of shear strain \exy from the surface of the piezo device to the top surface of the \SRO sample.
			As one could expect, the surface is the least strained part of the sample.
			For a sample thickness of \qty{30}{\micro\metre}, which is approximately the experimental thickness of all the \SRO samples used in this study, the simulated strain transfer is about \qty{80}{\percent}, which is consistent with the value of \qty{75}{\percent} experimentally measured in \cref{fig:Strain_transfer} of the main text.
			We can thus conclude that, for samples of thickness \qty{30}{\micro\metre}, the shear strain has only a small variation between \qtyrange{75}{100}{\percent} along the sample thickness.
		}
		\label{fig:ShearSimulation}
	\end{figure*}

	\begin{figure*}[htb]
		\includegraphics[page=7,width=100mm]{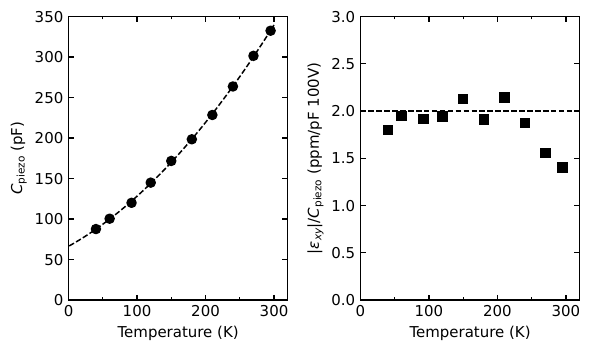}
		\subfloat{\label{fig:CapacitanceShear_vs_T}}
		\subfloat{\label{fig:ShearConversion_vs_T}}
		\caption{
			\textbf{Capacitance--strain conversion of a shear-piezo device.}
			\protect\subref{fig:CapacitanceShear_vs_T} Temperature dependence of the capacitance of a shear piezo device cut to the typical size of \qtyproduct{2x2x0.5}{\milli\metre} used in the experiments.
			The measurement is performed with a high-precision capacitance bridge (Andeen-Hagerling, AH 2500A) after subtracting the capacitance contribution from the cables.
			\protect\subref{fig:ShearConversion_vs_T} Ratio of the shear strain achieved at \qty{100}{\volt} and the piezo capacitance.
			The nearly constant value at lower temperatures allowed us to estimate a conversion factor $|\exy|/\Vp = \qty{0.007}{\percent\per 100 \volt}$ at \qty{2}{\kelvin}.
		}
		\label{fig:CapacitanceShearConversion}
	\end{figure*}
	
	\begin{figure}[htb]
		\includegraphics[page=8,width=130mm]{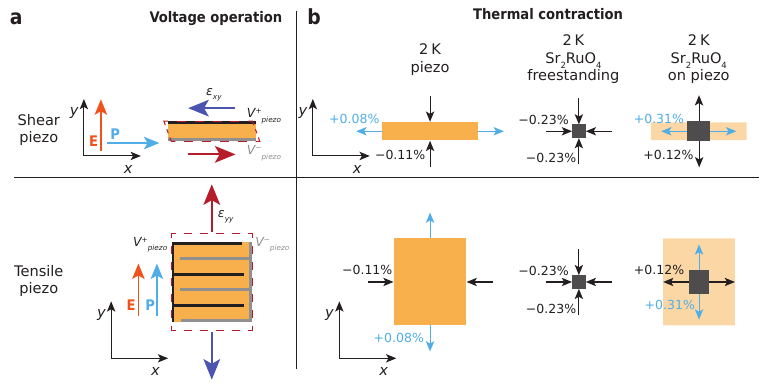}
		\subfloat{\label{fig:PreStrain_Voltage}}
		\subfloat{\label{fig:PreStrain_Thermal}}
		
		\caption{
			\textbf{Estimated pre-strains on \SRO due to differences in thermal contraction with piezo devices.}
			\protect\subref{fig:PreStrain_Voltage} The piezoelectric polarisation $P$ is perpendicular to the electric field $E$ in the case of a shear piezo (generating \exy with applied \Vp), while it is parallel to it in the case of a tensile piezo (generating uniaxial \eyy, or \exx with a different choice of coordinate system).
			The shear piezo used in this work is a single ceramic layer of \PZT (PZT), while the tensile piezo is a stack of PZT layers, with the electrodes indicated in black ($\Vp^+$) and grey ($\Vp^-$).
			\protect\subref{fig:PreStrain_Thermal}
			Upon cooling from room temperature down to \qty{2}{\kelvin}, the PZT material expands by \qty{+0.08}{\percent} in the direction of the polarisation and contracts by \qty{-0.11}{\percent} in the other directions~\cite{simpson1987thermal}.
			The \SRO material, in free-standing conditions, contracts by about \qty{-0.23}{\percent} in both $a$ and $b$ in-plane directions~\cite{chmaissem1998thermal}.
			Due to this difference in thermal contraction, the piezoelectric devices may determine a tensile pre-strain in \SRO up to \qty{+0.12} and \qty{+0.31}{\percent}, assuming a \qty{100}{\percent} strain transfer.
			In reality, the strain transferred is smaller due to the stiffness of \SRO and the deformation of the epoxy glue.
			This bi-axial pre-strain contains both compressive pre-strain and shear pre-strain components.
			In the case of the shear piezo, we expect that this pre-strain has no effect on our measurements because the voltage-induced strain lies in a symmetry channel different from the thermally induced strain.
			For example, when the shear piezo applies a voltage-induced strain \exy with \Bgg symmetry, the thermal pre-strain lies in the \Bg channel and is independent of \Vp.
			Therefore, by sweeping \Vp, we can apply both positive and negative shear strain and reliably traverse zero strain in the \Bgg channel.
			In the case of the tensile piezo, instead, the thermal pre-strain lies in the same symmetry channel as the voltage-induced strain, and thus shifts the zero-voltage strain value \cite{shapiro2016measurement}.
		}
		\label{fig:PreStrains}
	\end{figure}

	\begin{figure*}[htb]
		\includegraphics[page=9,width=120mm]{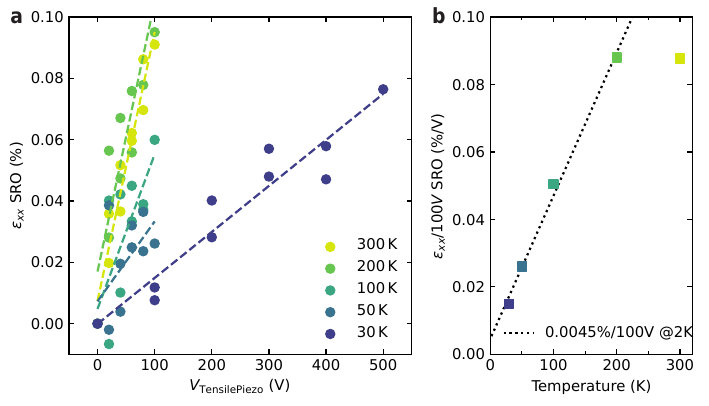}
		\subfloat{\label{fig:TensileCalibration_Curves}}
		\subfloat{\label{fig:TensileCalibration_Fit}}
		
		\caption{
			\textbf{Calibration of tensile strain on \SRO.}
			\protect\subref{fig:TensileCalibration_Curves} Uniaxial strain \exx measured by optical imaging of \SRO on the tensile piezo at different temperatures between 300 and \qty{30}{\kelvin}.
			The dashed lines are linear fits.
			\protect\subref{fig:TensileCalibration_Fit} Temperature dependence of the uniaxial strain achieved at \qty{100}{\volt}.
			The dotted line is a linear trend that is used to extrapolate the conversion factor of $\exx/\Vp = \qty{0.0045}{\percent\per 100 \volt}$ at \qty{2}{\kelvin}.
		}
		\label{fig:TensileCalibration}
	\end{figure*}
	
	\begin{figure*}[htb]
		\includegraphics[page=10,width=100mm]{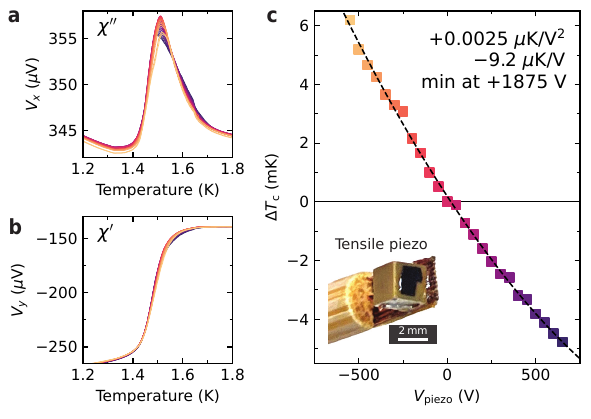}
		\subfloat{\label{fig:Tensile_Vx}}
		\subfloat{\label{fig:Tensile_Vy}}
		\subfloat{\label{fig:Tensile_vs_V}}
		
		\caption{
			\textbf{Extended data of \Tc variations with tensile strain in \SRO.}
			\protect\subref{fig:Tensile_Vx} Imaginary $\chi''$ and
			\protect\subref{fig:Tensile_Vy} real $\chi'$ part of the AC susceptibility measured by mutual inductance.
			The dissipation peak in $\chi''$ shifts towards higher temperatures upon application of a compressive strain ($\Vp<0$).
			\protect\subref{fig:Tensile_vs_V} Values of \Tc extracted from the mid-point of the drop in $\chi '$, with the fitting explained in the main text.
			Changes in \Tc of several milli-kelvin are detected with our high-precision method, validating our technique to detect small changes of \Tc due to applied strain.
		}
		\label{fig:TensileExtended}
	\end{figure*}
	
	\clearpage
	\section{Ginzburg-Landau theory for shear strain}
	\label{sec:GL}
	
	To understand the phase diagram of a two-component SC, we use a Ginzburg-Landau (GL) theory.
	Landau theory provides a simple model based on symmetry that can qualitatively describe OP behaviour near a phase transition.
	We use a GL free energy to derive the expected strain dependence of \Tc and compare the results of different experiments.
	
	The GL free energy is written as a power series expansion of an OP near a phase transition \cite{maeno2024still}.
	The free energy is an analytic function of the OP and it is invariant under the symmetries of the Hamiltonian.
	Because the free energy is a scalar quantity, it belongs to the \Ag irreducible representation (irrep).
	As an example, we consider the two-component OP $\bm{\eta} = (\eta_x, \eta_y)$ which belongs to the 2D \Eg irrep of the \Dfh point group with components $\{d_{xz}, d_{yz}\}$.
	Because the free energy is a scalar quantity and an analytic function of the OPs, the lowest order terms are quadratic in the OPs.
	The square of an \Eg order parameter has \Ag, \Bg, and \Bgg components; only combinations of the OP with \Ag symmetry are permitted in the free energy.
	Because we are interested in a minimal model that qualitatively describes the OP behaviour, we go only up to quartic order in the OP and consider the GL free energy density
	\begin{equation}
		\label{eq:GLenergy_SC}
		\begin{aligned}
			f_\eta &= a_0(T - \Tczero)|\bm{\eta}|^2 + \beta_1 |\bm{\eta}|^4
			+ \frac{\beta_2}{2} (\eta_x^2 \eta_y^{*2} + \eta_x^{*2} \eta_y^{2}) + \beta_3 |\eta_x|^2 |\eta_y|^2,
		\end{aligned}
	\end{equation}
	where $|\bm{\eta}|^2=|\eta_x|^2 + |\eta_y|^2$ and $a_0$ and $\beta_1$ are positive constants \cite{sigrist2002ehrenfest}.
	The $\beta_2$ term affects the relative phase between the SC order parameters, and thus governs whether the equilibrium state breaks time reversal symmetry. 
	For example, the chiral state $d_{xz}+id_{yz}$  breaks time reversal symmetry, whereas the nematic state $d_{xz} + d_{yz}$ does not.
	Both the $\beta_2$ and $\beta_3$ terms affect whether these SC order parameters coexist or compete.
	If the $\beta_3$ term is positive and has a larger magnitude than $\beta_2$, then the free energy is minimized by having one component equal to zero.
	For our analysis, we neglect spatial variation of the order parameter.
	
	The GL free energy density that describes the coupling between a two-component SC OP and lattice strain is given by
	\begin{equation}
		\label{eq:GLenergy}
		\begin{aligned}
			f_{\eta \varepsilon} &= r_1 \eAgxy|\bm{\eta}|^2 + r_2 \eAgz |\bm{\eta}|^2 
			+ r_3 \eBgg (\eta_x^* \eta_y + \eta_x \eta_y^*) + r_4 \eBg (|\eta_x|^2 - |\eta_y|^2),
		\end{aligned}
	\end{equation}
	where we go only to second order in $\eta_x$ and $\eta_y$ and first order in strain \cite{sigrist2002ehrenfest}. These strains, labelled in terms of their irreps, are given by
	\begin{equation}
		\begin{aligned}
			\eAgxy &= \exx + \varepsilon_{yy},  &&\eAgz = \varepsilon_{zz},\\
			\eBgg &= \varepsilon_{xy}, &&\eBg = \exx - \varepsilon_{yy}.
		\end{aligned}
	\end{equation}
	Here the strains are components of the linear strain tensor $\varepsilon_{ij}=(\partial u_i/\partial x_j + \partial u_j/\partial x_i)/2$ where $\bm{u}$ is a local displacement vector.
	Note that each free energy term has \Ag symmetry resulting from the combination of the strain and the OP-combination symmetries.
	There are no symmetry-allowed linear couplings to \Eg strains $\varepsilon_{xz}$ and $\varepsilon_{yz}$, so to lowest order these strains should not affect SC.
	The \Ag strains correspond to compressions or dilations of a sample that preserve the $C_{4z}$ rotational symmetry of the lattice; the \Bgg and \Bg strains correspond to volume-preserving shear distortions that break the $C_{4z}$ rotational symmetry of the lattice.
	
	We restrict our discussion here to the chiral \Eg SC phase to illustrate the coupling of a two-component OP to strain; \Eg is the only 2D even-parity irrep of the point group \Dfh. Other two-component OPs, discussed in literature, are composite of two different one-dimensional representations, such as the $d_{x^2 - y^2} + ig_{xy(x^2 - y^2)}$ state \cite{kivelson2020proposal, ghosh2021thermodynamic, maeno2024thirty}, combining \Bg for $d_{x^2 - y^2}$ and \Agg for $d_{xy(x^2 - y^2)}$, assumed to have nearly the same transition temperature.
	The advantage of considering the \Eg OP is that all examined strain channels can couple and allow us to derive the corresponding Ehrenfest relations as listed in Table 1 (up to some prefactors of order one). This is not the case for any of the composite two-component OPs. With this choice, we do not intend to promote any interpretation of the realized OP for this SC, as our experimental results are not supporting specifically a two-component order parameter. 
	\begin{figure*}[htb]
		\includegraphics[page=11,width=150mm]{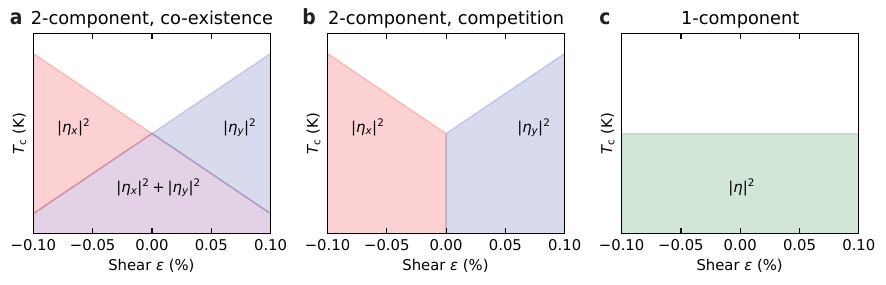}
		\subfloat{\label{fig:PhaseDiagram_2com_coex}}
		\subfloat{\label{fig:PhaseDiagram_2com_compet}}
		\subfloat{\label{fig:PhaseDiagram_1com}}
		
		\caption{
			\textbf{Phase diagrams from GL theory.}
			\protect\subref{fig:PhaseDiagram_2com_coex} Two-component SC when the GL free energy favours co-existence and \protect\subref{fig:PhaseDiagram_2com_compet} competition leading to a kink in the shear-strain dependence of \Tc.
			\protect\subref{fig:PhaseDiagram_1com}Example of a one-component SC, which is not affected by shear strain.
			All phase diagrams are only to linear order in strain; in principle, a higher-order quadratic strain dependence is allowed in all three scenarios. For two-component OPs, either \Bg or \Bgg shear strain can lead to a kink in \Tc vs strain, depending on the coupling constants $r_3$ and $r_4$ in \cref{eq:GLenergy}.
		}
		\label{fig:PhaseDiagram}
	\end{figure*}
	
	The total free energy density is given by the sum $f_{\text{total}} = f_\eta + f_{\eta \varepsilon}$.
	The GL transition temperatures are the temperatures at which the coefficients of $|\eta_x|^2$ and $|\eta_y|^2$ vanish.
	Assume for now that the quartic coefficients in \cref{eq:GLenergy_SC} favour coexistence of the two SC components.
	Because $|\eta_x|^2$ and $|\eta_y|^2$ have different coefficients for the \Bg and \Bgg shear strains, the SC components will have separate transition temperatures.
	Thus \Tc splitting is one of the experimental signatures of two-component SC \cite{sigrist1991phenomenological}, as we
	show in the example phase diagram of \cref{fig:PhaseDiagram_2com_coex}.
	If instead the quartic coefficients in \cref{eq:GLenergy_SC} favour competition, then only one of the SC components will become non-zero and shear strain will simply select which one is energetically favoured, much as a magnetic field selects whether spins point up or down for an Ising ferromagnet.
	A phase diagram for this scenario is shown in \cref{fig:PhaseDiagram_2com_compet}.
	In this case, compressive strain will linearly increase \Tc of one component, whereas tensile strain will linearly increase \Tc of the other component, and we expect a kink at the zero strain where there is a first order phase transition between the two OPs \cite{jerzembeck2024tc, hicks2025probing}.
	For a one-component SC, there is no linear coupling to shear strains, and thus there is no linear dependence of \Tc on strain as shown in \cref{fig:PhaseDiagram_1com}.
	
	We now consider the case of uniaxial strain.
	Due to the Poisson effect, upon applying strain \exx, there are also induced strains $\varepsilon_{yy}$ and $\varepsilon_{zz}$.
	Uniaxial strain \exx can therefore be decomposed into \Agxy, \Agz, and \Bg strains.
	For simplicity, let $a_0=1$.
	Then the GL transition temperatures for uniaxial strain are (to first order in strain) given by
	\begin{equation}
		\label{eq:exx}
		\begin{aligned}
			\Tcx &= \Tczero - r_1 \eAgxy - r_2 \eAgz - r_4 \eBg, \\
			\Tcy &= \Tczero - r_1 \eAgxy  - r_2  \eAgz + r_4  \eBg.
		\end{aligned}
	\end{equation}
	Note that the coupling constants can be interpreted as partial derivatives of \Tc with respect to their corresponding strains, e.g., $r_1 = -{\partial T_{\mathrm{c},k}}/{\partial \eAgxy}$ where $k = x$ or $y$. In magnetization or resistivity measurements, only the higher transition temperature $\Tc= \mathrm{max}(\Tcx, \Tcy)$ is observable. An experimental signature of two-component SC is the observation of a kink in $\Tc$ at $\exx = 0$ from the $r_4$ coupling, an effect which to our knowledge has not yet been observed \cite{hicks2014strong, steppke2017strong, watson2018micron, li2021high}. While \Tc splitting requires quartic coefficients that favour coexistence, a kink is expected to occur whether the SC components compete or coexist \cite{jerzembeck2024tc, hicks2025probing}.
	We comment that, since we are here interested in small applied strains, this analysis neglects higher-order quadratic strain couplings, which cause \Tc to depend quadratically on strain and may be responsible for the experimentally observed quadratic \Tc changes with larger \exx.
	
	Similarly, uniaxial strain \exxdiag applied along the diagonal $[110]$ axis can be decomposed into \Ag and \Bgg strains.
	To determine the effects of \Bgg shear strain, we can redefine the OPs as $\eta_x' = (\eta_x + \eta_y)/\sqrt{2}$ and $\eta_y' = (\eta_x - \eta_y)/\sqrt{2}$ to obtain the free energy density
	\begin{equation}
		\label{eq:GLenergy_B2g}
		\begin{aligned}
			f_{\eta' \varepsilon} &= r_1 \eAgxy|\bm{\eta'}|^2 + r_2 \eAgz |\bm{\eta'}|^2 
			+ r_4 \eBg (\eta_x'^* \eta_y' + \eta_x' \eta_y'^*) + r_3 \eBgg (|\eta_x'|^2 - |\eta_y'|^2).
		\end{aligned}
	\end{equation}
	By comparing \cref{eq:GLenergy_B2g} with \cref{eq:GLenergy}, we note that for \exxdiag the coefficients $r_3 \eBgg$ and $r_4 \eBg$ effectively switch roles, thus leading to
	\begin{equation}
		\begin{aligned}
			\Tcx' &= \Tczero - r_1 \eAgxy - r_2 \eAgz - r_3 \eBgg, \\
			\Tcy' &= \Tczero - r_1 \eAgxy  - r_2  \eAgz + r_3  \eBgg.
		\end{aligned}
	\end{equation}
	Similar to \cref{eq:exx}, these equations would lead to a kink at $\exxdiag=0$ with a slope given by the $r_3$ coupling constant.
	No kink in \Tc has yet been observed from \Bgg shear strain either \cite{hicks2014strong, jerzembeck2024tc}.
	It is possible that one or both of the coupling parameters $r_3$ and $r_4$ are small but non-zero, and hidden by larger \Ag terms as well as by higher-order quadratic strain couplings.
	These higher order couplings cause \Tc to depend on strain quadratically, and are observed in uniaxial \exx experiments \cite{hicks2014strong, steppke2017strong, watson2018micron, li2021high}.
	
	To circumvent this issue, we applied pure shear strains (\Bg or \Bgg) to search for kinks in \Tc. In the case of pure \Bg strain, the transition temperatures are given by
	\begin{equation}
		\begin{aligned}
			\Tcx  &= \Tczero - r_4 \eBg,  &\qquad\qquad \Tcx'  &= \Tczero - r_3 \eBgg, \\
			\Tcy  &= \Tczero + r_4 \eBg,  &\qquad\qquad \Tcy'  &= \Tczero + r_3 \eBgg.
		\end{aligned}
	\end{equation}
	Thus for both \Bg and \Bgg shear strains, we expect a linear strain dependence of \Tc with kinks at $\varepsilon=0$ for non-zero coupling parameters $r_3$ and $r_4$.
	Since our experiments of direct shear strain show no change in \Tc, we conclude that the parameters $r_3$ and $r_4$ must be close to zero, thus suggesting the absence of a two-component OP for \SRO.
	
	\subsection{1st Order Ehrenfest Relations}
	
	The coupling parameters $r_i$ can be indirectly estimated from ultrasound measurements using Ehrenfest relations \cite{sigrist2002ehrenfest}, which we show in \cref{tab:Ehrenfest} of the main text.
	The elastic free energy density is given by $f_\mathrm{el}=$$C_{ijkl}\varepsilon_{ij}\varepsilon_{kl}$/2, where $C_{ijkl}$ is the elasticity tensor and we use the Einstein summation convention \cite{chaikin1995principles}.
	For the \Dfh point group, this yields an elastic free energy density ~\cite{maeno2024still}
	\begin{equation}
		\begin{aligned}
			f_\mathrm{el} &= \frac{1}{2}\left[\frac{c_{11} + c_{12}}{2} \eAgxy^2 + c_{33}\eAgz^2 + 2c_{13}\eAgxy\eAgz
			+ \frac{c_{11} - c_{12}}{2} \eBg^2 + 4c_{66}\eBgg^2 + 4c_{44}(\ezx^2 + \eyz^2) \right],
		\end{aligned}
	\end{equation}
	where $c_{mn}$ are elastic moduli related to $C_{ijkl}$ by the following map for indices $m$ and $n: 1=xx$, $2=yy$, $3=zz$, $4=yz$, $5=xz$, and $6=xy$ \cite{chaikin1995principles}.
	By combining the GL free energy with the elastic free energy, it can be shown that there are jumps in elastic moduli at \Tc, and the jump magnitudes are proportional to the square of coupling parameters $r_i$ \cite{sigrist2002ehrenfest}.
	
	Ultrasound experiments report a jump in the $c_{66}$ elastic modulus \cite{
		benhabib2021ultrasound,%
		ghosh2021thermodynamic%
	}, although the size of the jump differs between the two experiments by two orders of magnitude, a discrepancy that in our view remains unresolved.
	Benhabib {\it et al.} \cite{benhabib2021ultrasound} used a pulse echo technique to measure the $c_{66}$ elastic modulus at \qty{169}{\mega\hertz} and found a jump magnitude of \qty{0.2}{ppm}; Ghosh {\it et al.} \cite{ghosh2021thermodynamic} used a resonant ultrasound technique to measure all the elastic moduli at lower frequencies on the order of \qty{2.5}{\mega\hertz}, and found a $c_{66}$ jump magnitude of \qty{17.5}{ppm}.
	We note that the pulse-echo technique tends to use strains of the order of \qtyrange{e-4}{e-6}{\percent}, which are larger than what used for the resonant ultrasound (strains of \qty{e-8}{\percent}) \cite{maeno2024thirty}.
	According to the supplement of Ref. \cite{ghosh2021thermodynamic}, larger frequencies may suppress the size of the $c_{ij}$ jumps at \Tc, but here we argue that the height of the jump should not be affected by the frequency.
	The frequency dependence of the $c_{66}$ jump is
	\begin{equation}
		\begin{aligned}
			\Delta c_{66} = \frac{r_3^2}{4\beta_2}\frac{1}{1 + \omega^2 \tau^2},
		\end{aligned}
	\end{equation}
	where $\tau$ is a relaxation time \cite{sigrist2002ehrenfest, ghosh2021thermodynamic}.
	Although it may appear that this equation shows that the jump is suppressed by higher frequencies, $\tau$ has a temperature dependence proportional to $(T - \Tc)^{-1}$ \cite{sigrist2002ehrenfest}, which diverges at \Tc, but is small away from the transition.
	Thus this equation shows that higher frequencies tend to broaden the jump rather than suppress the jump size.
	This fact is consistent with the experimental observation of a broader transition width in the higher frequency measurement of Ref. \cite{benhabib2021ultrasound}.
	
	Since the $c_{66}$ elastic modulus corresponds to the \Bgg mode, the observation of a jump in ultrasound experiments implies a non-zero $r_3$, suggesting that \SRO is a two-component SC.
	From Ref. \cite{sigrist2002ehrenfest}, for a chiral SC state (e.g., $d + id$), the \Bgg Ehrenfest relation  is given by
	\begin{equation}
		\label{eq:c66jump}
		\begin{aligned}
			\Delta c_{66} = -r_3^2\frac{\Delta C}{\Tc}\frac{4\beta_1 - \beta_2 + \beta_3}{8\beta_2},
		\end{aligned}
	\end{equation}
	where $\Delta C$ is the specific heat at constant volume jump at the SC transition. This Ehrenfest relation depends on several unknown quartic terms in the GL free energy; under the assumption of weak coupling (isotropic band approximation), we assume this total GL parameter factor is $1/2$ \cite{sigrist2002ehrenfest}, which allows us to calculate {\it order-of-magnitude} estimates.
	
	Finally, we discuss the specific heat at constant pressure $C_p$, which is related to the specific heat at constant volume $C_V$ by
	\begin{align}
		C_p = C_V +  T \frac{\alpha_p^2}{\kappa_T},
	\end{align}
	where $\alpha_p=(\partial V/ \partial T)/V$ is the volume thermal expansion coefficient at constant pressure, $\kappa_T=-(\partial V/ \partial p)/V$ is the volume compressibility at constant temperature, and the specific heat has units of \qty{}{\joule\per\metre\cubed\per\kelvin} \cite{chaikin1995principles}.
	Using the Debye temperature of \qty{410}{\kelvin} with 7 modes \cite{deguchi2004gap}, we estimate that $C_p$ for \SRO is about \qty{4.3e5}{\joule\per\metre\cubed\per\kelvin} at \qty{50}{\kelvin} .
	Near \qty{50}{\kelvin}, the factor $T \alpha_p^2/\kappa_T$ is about \qty{439}{\joule\per\metre\cubed\per\kelvin} \cite{chmaissem1998thermal}, or about 0.1\% of $C_p$.
	Since the thermal expansion $\alpha_p$ becomes smaller at lower temperatures \cite{chmaissem1998thermal}, $\Delta C_p \approx \Delta C_V$ is a good approximation near $\Tc = \qty{1.5}{\kelvin}$.
	
	From the data of Ref. \cite{
		ghosh2021thermodynamic%
	}, the Ehrenfest relation \cref{eq:c66jump} yields $|r_3| \sim \qty{750}{\milli\kelvin\per\percent}$;
	from the data of Ref. \cite{
		benhabib2021ultrasound%
	}, the Ehrenfest relation yields $|r_3| \sim \qty{70}{\milli\kelvin\per\percent}$.
	In Ref. \cite{
		ghosh2021thermodynamic%
	}, no jump was observed for the $c_{11} - c_{12}$ mode, which implies $r_4=0$ and there should be no \Tc splitting or kink from \Bg strain.
	
	\subsection{Stress Measurements}
	
	It is also possible to determine the coupling parameters $r_i$ by measuring the stress dependence of $\Tc$.
	In our measurements, we fixed the samples on piezoelectric devices and directly transmitted \textit{strains} to the sample.
	But in {\it stress} measurements, the sample is fixed in a free-standing configuration and is directly compressed along a crystallographic axis or through hydrostatic pressure.
	The stress tensor $\sigma_{ij}$ is given by the derivative of the elastic free energy density $f_\mathrm{el}$ with respect to the strain tensor $\varepsilon_{ij}$ \cite{chaikin1995principles}: $\sigma_{ij} = {\partial f_\mathrm{el}}/{\partial \varepsilon_{ij}} = C_{ijkl} \varepsilon_{kl}$. In vector form, this yields the following relation between applied stress $\vec{\sigma}$ and induced strains $\vec{\varepsilon}$:
	\begin{equation}
		\label{eq:stress_strain}
		\begin{bmatrix}
			\sigma_{xx} \\
			\sigma_{yy} \\
			\sigma_{zz} \\
			\sigma_{zx} \\
			\sigma_{yz} \\
			\sigma_{xy}
		\end{bmatrix}
		=
		\begin{bmatrix}
			c_{11}  & c_{12}  & c_{13}  & 0        & 0        & 0 \\
			c_{12}  & c_{11}  & c_{13}  & 0        & 0        & 0 \\
			c_{13}  & c_{13}  & c_{33}  & 0        & 0        & 0 \\
			0       & 0       & 0       & c_{44}  & 0        & 0 \\
			0       & 0       & 0       & 0        & c_{44}  & 0 \\
			0       & 0       & 0       & 0        & 0        & c_{66} \\
		\end{bmatrix}
		\begin{bmatrix}
			\exx \\
			\eyy \\
			\ezz \\
			2\ezx \\
			2\eyz \\
			2\exy
		\end{bmatrix}.
	\end{equation}
	By inverting \cref{eq:stress_strain}, one can determine the induced strains and use the GL free energy to relate the coupling constants $r_i$ to the measured stress dependence of $\Tc$. From hydrostatic pressure data in Ref. \cite{forsythe2002evolution}, we find $dT_\mathrm{c}^\mathrm{max}/d\ph = \qty{-216+-6}{\milli\kelvin\per\giga\pascal}$.
	From $c$-axis pressure data in Ref. \cite{jerzembeck2022superconductivity}, we find $dT_\mathrm{c}^\mathrm{max}/d\sigma_{zz} = \qty{79.0+-0.5}{\milli\kelvin\per\giga\pascal}$.
	Combining these derivatives with the elastic moduli from Ref. \cite{ghosh2021thermodynamic} yields $r_1=\qty{-33+-1}{\kelvin}$ and $r_2=\qty{-31.9+-0.5}{\kelvin}$.
	These results are smaller than the Ehrenfest relation estimates ($|r_1|=\qty{162}{\kelvin}$ and $|r_2|=\qty{168}{\kelvin}$) by about a factor of 5. References \cite{hicks2014strong, jerzembeck2024tc} both measured the response of \Tc with respect to stress $\sigma_{110}$, but did not observe any kink in \Tc indicating non-zero parameter $r_3$, in contrast to the ultrasound results; Ref. \cite{jerzembeck2024tc} placed an upper bound on $r_3$ of \qty{1.3}{\kelvin}, well below both Ehrenfest-relation estimates.

	\section{Basic definitions of elasticity}
	In the following, we provide some basic definitions that are useful when dealing with elasticity problems.
	The Voigt notation allows to transform the $3\times3$ stress and strain matrices (in three-dimensional space) of elements $\sigma_{ij}$, $\varepsilon_{ij}$ into $6\times1$ vectors of elements $\sigma_{k}$, $\varepsilon_{k}$ defined as
	\begin{equation}
		\begin{bmatrix}
			\sigma_1 \\
			\sigma_2 \\
			\sigma_3 \\
			\sigma_4 \\
			\sigma_5 \\
			\sigma_6
		\end{bmatrix}
		=
		\begin{bmatrix}
			\sigma_{xx} \\
			\sigma_{yy} \\
			\sigma_{zz} \\
			\sigma_{yz} \\
			\sigma_{zx} \\
			\sigma_{xy}
		\end{bmatrix},
		\qquad
		\begin{bmatrix}
			\varepsilon_1 \\
			\varepsilon_2 \\
			\varepsilon_3 \\
			\varepsilon_4 \\
			\varepsilon_5 \\
			\varepsilon_6
		\end{bmatrix}
		=
		\begin{bmatrix}
			\varepsilon_{xx} \\
			\varepsilon_{yy} \\
			\varepsilon_{zz} \\
			\gamma_{yz} \\
			\gamma_{zx} \\
			\gamma_{xy}
		\end{bmatrix}
		=
		\begin{bmatrix}
			\varepsilon_{xx} \\
			\varepsilon_{yy} \\
			\varepsilon_{zz} \\
			2\varepsilon_{yz} \\
			2\varepsilon_{zx} \\
			2\varepsilon_{xy}
		\end{bmatrix},
	\end{equation}
	where the factors 2 on the shear components are necessary to preserve the calculation of the elastic energy density $u$ as product of $\sigma$ and $\varepsilon$ which, using the Einstein summation convention on repeated indexes, reads
	\begin{equation}
		\begin{aligned}
			u = \frac{1}{2} \sigma_{ij}\varepsilon_{ij}
			&= \frac{1}{2}\left( 
			\sigma_{xx} \varepsilon_{xx}
			+ \sigma_{yy} \varepsilon_{yy}
			+ \sigma_{zz} \varepsilon_{zz}
			+ \sigma_{xy} \varepsilon_{xy}
			+ \sigma_{yx} \varepsilon_{yx}
			+ \sigma_{xz} \varepsilon_{xz}
			+ \sigma_{zx} \varepsilon_{zx}
			+ \sigma_{yz} \varepsilon_{yz}
			+ \sigma_{zy} \varepsilon_{zy}
			\right)
			\\
			&= \frac{1}{2}\left(\sigma_{xx} \varepsilon_{xx}
			+ \sigma_{yy} \varepsilon_{yy}
			+ \sigma_{zz} \varepsilon_{zz}
			+ 2\sigma_{xy} \varepsilon_{xy}
			+ 2\sigma_{xz} \varepsilon_{xz}
			+ 2\sigma_{yz} \varepsilon_{yz}
			\right),
		\end{aligned}
	\end{equation}
	and equivalently in Voigt notation
	\begin{equation}
		u=\frac{1}{2}\sigma_i\varepsilon_i =
		\frac{1}{2}\left(
		\sigma_1 \varepsilon_1 +
		\sigma_2 \varepsilon_2 +
		\sigma_3 \varepsilon_3 +
		\sigma_4 \varepsilon_4 +
		\sigma_5 \varepsilon_5 +
		\sigma_6 \varepsilon_6
		\right).
	\end{equation}
	In Voigt notation, the $3\times3\times3\times3$ elasticity tensor $c_{ijkl}$ is reduced to a $6\times6$ matrix of elastic moduli $c_{ij}$ which is symmetric and defined so that
	\begin{equation}
		\begin{bmatrix}
			\sigma_1 \\
			\sigma_2 \\
			\sigma_3 \\
			\sigma_4 \\
			\sigma_5 \\
			\sigma_6
		\end{bmatrix}
		=
		\begin{bmatrix}
			c_{11} & c_{12} & c_{13} & c_{14} & c_{15} & c_{16} \\
			c_{21} & c_{22} & c_{23} & c_{24} & c_{25} & c_{26} \\
			c_{31} & c_{32} & c_{33} & c_{34} & c_{35} & c_{36} \\
			c_{41} & c_{42} & c_{43} & c_{44} & c_{45} & c_{46} \\
			c_{51} & c_{52} & c_{53} & c_{54} & c_{55} & c_{56} \\
			c_{61} & c_{62} & c_{63} & c_{64} & c_{65} & c_{66}
		\end{bmatrix}
		\begin{bmatrix}
			\varepsilon_1 \\
			\varepsilon_2 \\
			\varepsilon_3 \\
			\varepsilon_4 \\
			\varepsilon_5 \\
			\varepsilon_6
		\end{bmatrix}.
	\end{equation}
	In the case of \SRO, in the point group \Dfh, only 6 coefficients are necessary to fully describe the stress--strain relationship (the symmetric part of the matrix is omitted for simplicity of display)
	\begin{equation}
		\begin{bmatrix}
			\sigma_{xx} \\
			\sigma_{yy} \\
			\sigma_{zz} \\
			\sigma_{yz} \\
			\sigma_{zx} \\
			\sigma_{xy}
		\end{bmatrix}
		=
		\begin{bmatrix}
			c_{11}&{\color{Orange}c_{12}}&{\color{Purple}c_{13}}&{\color{Gray}0}&{\color{Gray}0}&{\color{Gray}0}\\
			&c_{11}&{\color{Purple}c_{13}}&{\color{Gray}0}&{\color{Gray}0}&{\color{Gray}0}\\
			&&{\color{red}c_{33}}&{\color{Gray}0}&{\color{Gray}0}&{\color{Gray}0}\\
			&&&{\color{Blue}c_{44}}&{\color{Gray}0}&{\color{Gray}0}\\
			&&&&{\color{Blue}c_{44}}&{\color{Gray}0}\\
			&&&&&{\color{Green}c_{66}}\\
		\end{bmatrix}
		\begin{bmatrix}
			\varepsilon_{xx} \\
			\varepsilon_{yy} \\
			\varepsilon_{zz} \\
			2\varepsilon_{yz} \\
			2\varepsilon_{zx} \\
			2\varepsilon_{xy}
		\end{bmatrix}.
	\end{equation}

	\noindent
	It is also useful to define the irrep \Agxy and \Bg stress and strains as
	\begin{align}
		\sAgxy &= \frac{\sigma_{xx} + \sigma_{yy}}{2} = \frac{\sigma_1 + \sigma_2}{2},
		&\quad
		\eAgxy &= \varepsilon_{xx} + \varepsilon_{yy} = \varepsilon_1 + \varepsilon_2, \\
		\sBg &= \frac{\sigma_{xx} - \sigma_{yy}}{2} = \frac{\sigma_1 - \sigma_2}{2},
		&
		\eBg &= \varepsilon_{xx} - \varepsilon_{yy} = \varepsilon_1 - \varepsilon_2,\\
		\sigma_{1,2} &= \sAgxy \pm \sBg,
		&
		\varepsilon_{1,2} &= \frac{\eAgxy \pm \eBg}{2},
	\end{align}
	where the factor $1/2$ is usually associated either to the stress (as we do here) or to the strain.

	\section{Derivation of \texorpdfstring{\cref{eq:xy_from_ThreeExperiments}}{Eq. (3)} in the main text by combination of three stress--strain experiments}
	To calculate the changes in \Tc of \SRO, we now consider the derivatives with respect to stress and strain.
	For the \Agxy and \Bg, using the chain rule, we have
	\begin{align}
		\label{eq:dTc_dsAg}
		\frac{\partial \Tc}{\partial \sAgxy} &= 
		\frac{\partial \Tc}{\partial \sigma_i} \frac{\partial \sigma_i}{\partial \sAgxy} = 
		\frac{\partial \Tc}{\partial \sigma_{xx}} + \frac{\partial \Tc}{\partial \sigma_{yy}},
		\\
		\label{eq:dTc_dsBg}
		\frac{\partial \Tc}{\partial \sBg} &= 
		\frac{\partial \Tc}{\partial \sigma_i} \frac{\partial \sigma_i}{\partial \sBg} = 
		\frac{\partial \Tc}{\partial \sigma_{xx}} - \frac{\partial \Tc}{\partial \sigma_{yy}}.
	\end{align}

	\noindent
	In the case of hydrostatic pressure, which is defined with opposite sign to stresses, we have
	$d\ph= -d\sigma_i$, with $i=1,2,3$ (the other components are zero) and calculate the total differential of stress as
	\begin{equation}
		d\Tc = \frac{\partial \Tc}{\partial \sigma_i}d\sigma_i=
		-\left( \frac{\partial \Tc}{\partial \sigma_{xx}} + \frac{\partial \Tc}{\partial \sigma_{yy}} + \frac{\partial \Tc}{\partial \sigma_{zz}} \right) d\ph,
	\end{equation}
	from which we obtain
	\begin{equation}
		\label{eq:Agxy_from_phydro}
		\frac{\partial \Tc}{\partial \sAgxy} = -\frac{d\Tc}{d\ph} -\frac{\partial \Tc}{\partial \sigma_{zz}}.
	\end{equation}
	
	\noindent
	To calculate the effect of stress along the [110] in-plane direction, it is useful to consider the three-dimensional rotation matrices.
	We can thus consider a uniaxial stress of the kind \sxx with magnitude $\sigma^*$ applied to a reference system rotated by \qty{-45}{\degree} along the $z$ axis and then rotated back by \qty{+45}{\degree} as
	\begin{equation}
		\begin{aligned}
			\sxxdiag = R_z(+45^\circ)\sigma_{xx}R_z(-45^\circ)
			=
			\frac{1}{2}
			\begin{bmatrix}
				1 & -1 & 0 \\
				1 &  1 & 0 \\
				0           &  0           & \sqrt{2} \\
			\end{bmatrix}
			\begin{bmatrix}
				\sigma^* & 0 & 0 \\
				0 &  0 & 0 \\
				0           &  0           & 0 \\
			\end{bmatrix}
			\begin{bmatrix}
				1 & 1 & 0 \\
				- 1 &  1 & 0 \\
				0           &  0           & \sqrt{2} \\
			\end{bmatrix}
			=
			\frac{1}{2}
			\begin{bmatrix}
				\sigma^* & \sigma^* & 0 \\
				\sigma^* &  \sigma^* & 0 \\
				0           &  0           & 0 \\
			\end{bmatrix}.
		\end{aligned}
	\end{equation}
	Thus, the uniaxial stress \sxxdiag involves 
	\begin{align}
		\sigma_{xx} = \sigma_{yy} = \sigma_{xy} = \frac{\sigma^*}{2}.
	\end{align}
	We can calculate the total change of \Tc as
	\begin{align}
		d\Tc = \frac{\partial \Tc}{\partial \sigma_i}d\sigma_i
		=
		\left( \frac{\partial \Tc}{\partial \sigma_{xx}} + \frac{\partial \Tc}{\partial \sigma_{yy}} + \frac{\partial \Tc}{\partial \sigma_{xy}} \right) \frac{1}{2} d\sxxdiag,
	\end{align}
	and then apply \cref{eq:dTc_dsAg} to obtain the final result
	\begin{align}
		\label{eq:sdiag_from_Agxy}
		\frac{d \Tc}{d \sxxdiag} =
		\frac{1}{2} \left( \frac{\partial \Tc}{\partial \sAgxy} + \frac{\partial \Tc}{\partial \sigma_{xy}} \right).
	\end{align}
	
	\noindent
	Similarly, we can calculate the effect of $xy$ shear strain of magnitude $\sigma^*$ along the [110] in-plane direction as
	\begin{equation}
		\begin{aligned}
			\sxydiag &= R_z(+45^\circ)\sigma_{xy}R_z(-45^\circ)
			=
			\frac{1}{2}
			\begin{bmatrix}
				1 & -1 & 0 \\
				1 &  1 & 0 \\
				0           &  0           & \sqrt{2} \\
			\end{bmatrix}
			\begin{bmatrix}
				0 & \sigma^* & 0 \\
				\sigma^* &  0 & 0 \\
				0           &  0           & 0 \\
			\end{bmatrix}
			\begin{bmatrix}
				1 & 1 & 0 \\
				- 1 &  1 & 0 \\
				0           &  0           & \sqrt{2} \\
			\end{bmatrix}
			=
			\frac{1}{2}
			\begin{bmatrix}
				-2\sigma^* & 0 & 0 \\
				0 &  2\sigma^* & 0 \\
				0           &  0           & 0 \\
			\end{bmatrix},
		\end{aligned}
	\end{equation}
	which involves
	\begin{equation}
		\sigma_{xx} = -\sigma^*,\quad
		\sigma_{yy} = +\sigma^*,
	\end{equation}
	leading to
	\begin{align}
		d\Tc = \frac{\partial \Tc}{\partial \sigma_i}d\sigma_i
		=
		\left( -\frac{\partial \Tc}{\partial \sigma_{xx}} + \frac{\partial \Tc}{\partial \sigma_{yy}} + \right) d\sxydiag.
	\end{align}
	By making use of \cref{eq:dTc_dsBg}, we get the final result
	\begin{align}
		\frac{d \Tc}{d \sxydiag} =
		-\frac{\partial \Tc}{\partial \sigma_\Bg},
	\end{align}
	which shows that \sxydiag shear stress is equivalent to pure \Bg stress, except for the minus sign.
	We report these results, along with stress--strain conversions obtained by making use of the chain rule, in \cref{tab:StressStrain}.
	Finally, combining \cref{eq:Agxy_from_phydro,eq:sdiag_from_Agxy} we obtain \cref{eq:xy_from_ThreeExperiments} in the main text that is used to predict the dependence of \Tc on shear strain $\sigma_{xy}$ from results of hydrostatic pressure, uniaxial stress along the $z$ and the [110] directions.

	\begin{table*}[htb]
		\centering
		\caption{\textbf{Relationships of \Tc variations extracted by different experiments of uniaxial stress and hydrostatic pressure.}
			The relationships are obtained as described in this section and are used to calculate \cref{eq:xy_from_ThreeExperiments} in the main text.
			For clarity, the results in the table are provided by replacing the Voigt notation with the explicit $x$,$y$,$z$ indexes of spatial coordinates.}
		\label{tab:StressStrain}
		\begin{tabular}{r|c|c}
			& %
			Stress &
			Strain\\
			\midrule
			
			Shear $xy$ &
			$\displaystyle
			\frac{\partial \Tc}{\partial \sigma_{xy}} =
			2\frac{d\Tc}{d\sxxdiag} -\frac{\partial \Tc}{\partial \sAgxy}
			$
			&
			$\displaystyle
			\frac{\partial \Tc}{\partial \varepsilon_{xy}} = 
			2c_{66}\frac{\partial \Tc}{\partial \sigma_{xy}}
			$
			\\[4mm]
			
			Shear \Bg &
			$\displaystyle
			\frac{d\Tc}{d\sxydiag} =
			-\frac{\partial\Tc}{\partial\sBg}
			$
			&
			$\displaystyle
			\frac{\partial \Tc}{\partial \eBg} =
			\frac{c_{11} - c_{12}}{2}\frac{\partial \Tc}{\partial \sBg}
			$
			\\[4mm]
			
			Shear $yz$, $zx$ &
			---&%
			$\displaystyle
			\frac{\partial \Tc}{\partial \varepsilon_{yz}} = 
			2c_{44}\frac{\partial \Tc}{\partial \sigma_{yz}}
			$
			\\[4mm]
			
			\midrule
			
			Biaxial compressive \Agxy &
			$\displaystyle
			\frac{\partial \Tc}{\partial \sAgxy} =
			-\frac{d\Tc}{d\ph} -\frac{\partial \Tc}{\partial \sigma_{zz}}
			$
			&
			$\displaystyle
			\frac{\partial \Tc}{\partial \eAgxy} = 
			\frac{c_{11} + c_{12}}{2}\frac{\partial \Tc}{\partial \sAgxy} +
			c_{13}\frac{\partial \Tc}{\partial \sigma_{zz}}
			$
			\\[4mm]
			
			Compressive $zz$ &
			---& %
			$\displaystyle
			\frac{\partial \Tc}{\partial \varepsilon_{zz}} = 
			c_{13} \frac{\partial \Tc}{\partial \sAgxy}+
			c_{33}\frac{\partial \Tc}{\partial \sigma_{zz}}
			$
			\\[4mm]
			
		\end{tabular}
	\end{table*}

\end{document}